\documentclass[aps,prb,preprint,groupedaddress,floatfix]{revtex4}
\usepackage{graphicx,amssymb,array,amsmath}
\begin{document}
\title{Entropy determination for mixtures in the adiabatic grand-isobaric ensemble}
\author{Caroline Desgranges$^4$ and Jerome Delhommelle$^{1,2,3,4}$}
\affiliation{$^1$ Department of Chemistry, University of North Dakota}
\affiliation{$^2$ Department of Biomedical Engineering, University of North Dakota}
\affiliation{$^3$ School of Electrical Engineering and Computer Science, University of North Dakota}
\affiliation{$^4$ MetaSimulation of Nonequilibrium Processes (MSNEP), Tech Accelerator, University of North Dakota}

\date{\today}

\begin{abstract}
The entropy change that occurs upon mixing two fluids has remained an intriguing topic since the dawn of statistical mechanics. In this work, we generalize the grand-isobaric ensemble to mixtures, and develop a Monte Carlo algorithm for the rapid determination of entropy in these systems. A key advantage of adiabatic ensembles is the direct connection they provide with entropy. Here, we show how the entropy of a binary mixture A-B can be readily obtained in the adiabatic grand-isobaric $(\mu_{\text{A}}$, $\mu_{\text{B}}, P, R)$ ensemble, in which $\mu_{\text{A}}$ and $\mu_{\text{B}}$ denote the chemical potential of components A and B, respectively, $P$ is the pressure, and $R$ is the heat (Ray) function, that corresponds to the total energy of the system. This, in turn, allows for the evaluation of the entropy of mixing, as well as of the Gibbs free energy of mixing. We also demonstrate that our approach performs very well both on systems modeled with simple potentials and with complex many-body force fields. Finally, this approach provides a direct route to the determination of the thermodynamic properties of mixing, and allows for the efficient detection of departures from ideal behavior in mixtures.
\end{abstract}

\maketitle

\section{Introduction}

Entropy has intrigued thermodynamicists for centuries, and a full understanding of this concept still remains elusive. Among other processes, mixing, and the associated entropy of mixing, has challenged scientists since the inception of statistical mechanics, from Boltzmann's early definition of entropy to the proposal of the Gibbs paradox. Several studies have focused on the determination of the entropy of mixing for liquid mixtures over the years, both experimentally and theoretically~\cite{frank1945free,wood1947entropy,brown1957statistical,brown1957statistical2,flory1965statistical,abe1965thermodynamic,lacombe1976statistical,hoshino1980entropy,panayiotou1984two,tanaka1990relationship,sommer2001entropy,rowlinson2013liquids,arzpeyma2013prediction}. In binary mixtures of molecular compounds, three factors are found to contribute to the excess entropy of mixing: (i) the relative volumes of the molecules, (ii) the spatial distribution of molecules about a reference molecule, and (iii) the non-random orientational distribution of molecule about a reference molecule~\cite{wood1947entropy}. On the other hand, in binary mixtures of atomic fluids, the excess entropy of mixing only depends on the free volume of each component. This has given rise to the concept of the combinatorial entropy of mixing~\cite{huggins1971thermodynamic,lichtenthaler1973combinatorial,donohue1975combinatorial}, which stems from the randomness in placing the atoms in the system's volume. Then, the theoretical entropy of mixing in excess of the combinatorial entropy, or ideal entropy, is obtained from the reduced volumes of the two mixture components. Other approaches include conventional solution theories, which have relied exclusively on two features of the liquid mixtures. The first is the entropy of dispersion of the two molecular species, generally evaluated using a lattice model~\cite{prigogine1953application,prigogine1956statistical}. The second comes from the interactions between neighboring molecules and is found to depend on the difference in interactions between pairs of like and unlike molecules. This approach has led to excellent results for the synthesis and processing of high-entropy alloys~\cite{kube2019phase,marshal2017combinatorial,li2018combinatorial,shi2020high}. It has also been shown recently that the Shannon entropy can quantify the amount of disorder within a system. In mixtures, the entropy of mixing is defined as the increase in disorder upon the transition from a fully demixed state to an ideally mixed state~\cite{camesasca2006quantifying,brandani2013quantifying}. This approach has enabled to quantify the quality of mixing in polymers, and to design, control and asses optimal mixing protocols~\cite{d1999control,camesasca2006quantifying}. The Shannon or information entropy can also be calculated through the pair correlation functions~\cite{sastry2000evaluation,donev2007configurational,qin2012new,nicholson2021entropy}. Such an approach has been used recently to predict the entropy of liquid aluminum, copper and aluminum-copper alloys~\cite{gao2018information}.

Here, we propose a different approach for the determination of the entropy of mixing. To this end, we develop a method based on the adiabatic thermodynamics formalism to derive a direct route to the entropy. This approach has been developed so far for single-component systems by Ray {\it et al.}~\cite{ray1981new, ray1993monte,ray1993new, graben1991unified, ray1990fourth, ray1986fundamental, ray1996microcanonical, ray1991microcanonical}. For a binary mixture A-B, this leads to working in an ensemble for which ($\mu_{\text{A}},\mu_{\text{B}}, P, R$) are fixed. In this set, $\mu_{\text{A}}$ denotes the chemical potential for the component A, $\mu_{\text{B}}$ the chemical potential for component B, $P$ the pressure, and $R$ the heat or Ray function. We can then use the simple relationship $S=R/T$ to gain access to the entropy of the system. Another advantage of this ensemble is that pressure is an input parameter, and that its calculation via, for instance, the virial expression is not needed in the course of the simulations. This is especially interesting for systems modeled with complex many-body potentials. We add that the determination of temperature is straightforward in this adiabatic ensemble, as it stems from the equipartition principle. This provides a simple and versatile framework and method to compute the entropy of mixing.

The paper is organized as follows. We first present the generalization of the adiabatic grand-isobaric ensemble to the case of mixtures, the implementation of Monte Carlo simulations in this ensemble, as well as the models used in this work to model mixtures of simple fluids and binary mixtures of liquid metals. We then discuss the results obtained in the grand-isobaric ensemble for the Neon-Argon mixture, as well as copper-silver. In both cases, we present results for the thermodynamic properties of mixing, including the entropy, as well as the enthalpy and the Gibbs free energy of mixing. To assess accuracy, we compare the results to the available experimental data and to results obtained with conventional simulation methods in isothermal ensembles. We finally draw the main conclusions from this work in the last section.

\section{Formalism and simulation methods}

\subsection{Adiabatic ensemble framework}

We start with a brief discussion of how the probability density, and thus the acceptance probabilities used in Monte Carlo simulations, are obtained for adiabatic ensembles in the case of single-component systems. In the microcanonical ensemble, the number of accessible microstates~\cite{naudts2005boltzmann, ray1999correct, pearson1985laplace, turban2013recursive} is given by $\Omega (N,V,E)= \rho(E) \delta E$, in which $\rho(E)$ is the microcanonical density of states and $\delta E << E$. To show how the probability density can be derived, we start with the case of an ideal gas of $N$ particles of mass $m$ in a volume $V$, we have the following Hamiltonian
\begin{equation}
H= \sum_{i=1}^{3N} \frac{\mathbf{p_i}^2}{2m}=K 
\end{equation}
in which $\mathbf{p}_i$ denotes the momentum of particle $i$ and $K$ the total kinetic energy for the system. The phase space volume $\Omega$ can be defined as
\begin{equation}
\begin{array}{lll}
\Omega & = & \int...\int dq_{3N} \int ...\int dp_1....dp_{3N} 
\end{array}
\label{Omega}
\end{equation}
for sets of coordinates and momenta such that $0 \leq H(\mathbf{q}_i,\mathbf{p}_i) \leq E$.
This gives, after integration over the position coordinates
\begin{equation}
\begin{array}{lll}
\Omega & = & V^N \int ...\int dp_1....dp_{3N}
\end{array}
\label{OmegaIG}
\end{equation}
for sets of momenta such that $0 \leq \sum_{i=1}^{3N} \frac{p_i^2}{2m}\leq E$. As discussed in prior work~\cite{fernandez1979dirichlet}, Dirichlet's integral formula can be used to calculate this integral. We recall that Dirichlet's integral formula states that
\begin{equation}
\begin{array}{ccc}
I & = & \int ...\int t_1^{\alpha_1 -1} t_2^{\alpha_2 -1}...t_n^{\alpha_n -1} dt_1 dt_2...dt_n \\
 & = & \frac{b_1^{\alpha_1} b_2^{\alpha_2}...b_n^{\alpha_n}}{\beta_1 \beta_2...\beta_n}\times \frac{\Gamma(\alpha_1/\beta_1)\Gamma(\alpha_2/\beta_2)...\Gamma(\alpha_n/\beta_n)}{\Gamma (\alpha_1/\beta_1 + \alpha_2/\beta_2+...+\alpha_n/\beta_n +1)}\\
\end{array}
\label{resDir}
\end{equation}
in which $t_i,b_i,\beta_i$ are positive and such that
\begin{equation}
(t_1/b_1)^{\beta_1}+(t_2/b_2)^{\beta_2}+...+(t_n/b_n)^{\beta_n} \leq 1
\end{equation}
To solve Eq.~\ref{OmegaIG}, we use Eq.~\ref{resDir} with $\alpha_i=1, t_i=p_i, \beta_i=2, b_i=(2mE)^{1/2}$, (i=1,2,...,3N) to obtain
\begin{equation}
\Omega=\frac{{2^{3N}V^N (2mE)}^{3N/2} [\Gamma(1/2)]^{3N}}{2^{3N}\Gamma(3N/2+1)}
\label{Omf}
\end{equation}
and account for the fact that particles are indistinguishable, and that the phase space volume is dimensionless, by dividing Eq.~\ref{Omf} by $N!$ and $h^{3N}$
\begin{equation}
\Omega=\frac{V^N}{N!} \left( {\frac{2 \pi m}{h^2}} \right)^{3N/2} \frac{E^{3N/2}}{\Gamma(3N/2+1)}
\end{equation} 
in which we use $\Gamma(1/2)=\sqrt \pi$. This yields the density of states $\rho(E)$ by differentiating the phase space volume with respect to $E$ as
\begin{equation}
\rho(E)=\frac{\partial \Omega}{\partial E}=\frac{V^N}{N!} \left( {\frac{2 \pi m}{h^2}} \right)^{3N/2} \frac{E^{3N/2-1}}{\Gamma(3N/2)}
\end{equation}

As shown by Ray {\it et al.}~\cite{ray1991microcanonical,ray1993monte}, this formalism can be generalized to other types of systems (such as, {\it e.g.}, with Hamiltonian of the form $H=K+U(\mathbf{q})$), in which $U(\mathbf{q})$ denotes a position-dependent potential energy) in the microcanonical ensemble, and to other adibabatic ensembles (such as, {\it e.g.}, the adiabatic grand-isobaric ensemble $(\mu, P, R)$). The key here is to realize that the $K=E$ for an ideal gas in a microcanonical ensemble becomes either $K=E-U(\mathbf{q})$ for an interacting system in the microcanonical ensemble, or $K=R-PV+\mu N -U(\mathbf{q})$ for an interacting system in the adiabatic grand-isobaric ensemble. The $(\mu, P, R)$ ensemble models an open adiabatically insulated system in contact with a pressure reservoir and a chemical potential reservoir. Here the energy $R$ is related to the enthalpy $H$ by $R=H-\mu N$.

We can then determine the acceptance probability $acc(o \rightarrow n)$ for the microcanonical Monte Carlo method. Here a MC move is attempted from an ``old'' configuration $o$ with a set of positions denoted by $\mathbf{q}$ to a ``new'' configuration $n$ with a set of positions denoted by $\mathbf{q'}$ yielding the following acceptance probability
\begin{equation}
\begin{array}{lll}
acc(o \rightarrow n) & = & \min \left[ 1,\frac{\rho (\mathbf{q'}, N, V)}{\rho (\mathbf{q}, N, V)} \right]\\
 & = & \min \left[ 1,\frac{K_n^{3N/2-1}}{K_o^{3N/2-1}} \right]\\
 & = & \min \left[ 1,\frac{(E-U(\mathbf{q}'))^{3N/2-1}}{(E-U(\mathbf{q}))^{3N/2-1}} \right]\\
\end{array}
\end{equation}

Similarly, in the adiabatic grand-isobaric ensemble~\cite{ray1993monte,desgranges2020central}, if the ``old'' configuration is denoted by $(\mathbf{q}, N, V)$ and the ``new'' configuration by $(\mathbf{q}', N', V')$, the acceptance probability can be written as
\begin{equation}
\begin{array}{lll}
acc(o \rightarrow n) & = & \min \left[ 1, \frac{\rho(\mathbf{q}',N',V')}{\rho(\mathbf{q},N,V)}\right] \\
& = & \min\left[ {1, \frac{ (bV')^{N'} N! \Gamma(3N/2) K_n^{3N'/2-1}}{(bV)^{N} N'! \Gamma(3N'/2) K_o^{3N/2-1}}} \right]\\
& = & \min\left[ {1, \frac{ (bV')^{N'} N! \Gamma(3N/2) (R-PV'+\mu N'-U(\mathbf{q}'))^{3N'/2-1}}{(bV)^{N} N'! \Gamma(3N'/2) (R-PV+\mu N-U(\mathbf{q}))^{3N/2-1}}} \right]\\
\end{array}
\label{genmuPR}
\end{equation}
in which $b=(2\pi m / h^2)^{3/2}$. From a practical standpoint and to ensure high acceptance probabilities, MC moves are split into 4 different categories, corresponding to translations, insertions, deletions and volume changes. We extend the formalism to the case of mixtures, and discuss in greater detail these acceptance probabilities for 2-component systems, in the next section.

\subsection{Adiabatic formalism for multi-component systems}

In this section, we extend the adiabatic framework and the adiabatic grand-isobaric ensemble to systems with multiple components. In line with the derivation for single component systems, we start with a mixture of ideal gases and determine the phase space volume and probability density. For a mixture A-B of two ideal gases A and B, we have the following Hamiltonian
 \begin{equation}
 H=\sum_{i=1}^{3N_{\text{A}}} \frac{p_i^2}{2m_{\text{A}}}+\sum_{j=1}^{3N_{\text{B}}} \frac{p_j^2}{2m_{\text{B}}}=K
 \end{equation}
 in which $N_{\text{A}}$ and $N_{\text{B}}$ denote the number of particles for the two components A and B, and $m_{\text{A}}$ and $m_{\text{B}}$ their respective masses.
 
 The phase space volume is given by
 \begin{equation}
 \Omega=V^{N_{\text{A}}+N_{\text{B}}} \int...\int \prod_i^{3N_{\text{A}}} dp_i \prod_j^{3N_{\text{B}}} dp_j
 \end{equation}
 
 As with single-component systems, we now use Dirichlet's integral formula to calculate $\Omega$. In the case of the binary mixture A-B, we use Eq.~\ref{resDir} with the following parameters: $\alpha_k=1$, $t_k=p_i$, $\beta_k=2$, $b_k=(2m_{\text{A}}E)^{1/2}$, $k=1,2,...,3N_{\text{A}}$, $\alpha_k=1$, $t_k=p_j$, $\beta_k=2$, $b_k=(2m_{\text{B}}E)^{1/2}$, $k=3N_A+1,...,3N_{\text{A}}+3N_{\text{B}}$, and carry out the integration over the following domain
 \begin{equation}
 \sum_{i=1}^{3N_{\text{A}}} \left( {\frac{p_i}{(2m_{\text{A}}E)^{1/2}}}\right)^2 + \sum_{j=1}^{3N_{\text{B}}} \left( {\frac{p_j}{(2m_{\text{B}}E)^{1/2}}}\right)^2 \le 1
 \end{equation}
 This yields the following result for the phase space volume
 \begin{equation}
 \Omega= \frac{1}{h^{3(N_{\text{A}}+N_{\text{B}})}N_{\text{A}}!N_{\text{B}}!}V^{N_{\text{A}}+N_{\text{B}}}\frac{(2\pi m_{\text{A}})^{3N_{\text{A}}/2}(2\pi m_{\text{B}})^{3N_{\text{B}}/2}}{\Gamma[3(N_{\text{A}}+N_{\text{B}})/2+1]} E^{3(N_{\text{A}}+N_{\text{B}})/2}
 \end{equation}
 and for the probability density
 \begin{equation}
 \rho(E)= \frac{1}{h^{3(N_{\text{A}}+N_{\text{B}})}N_{\text{A}}!N_{\text{B}}!}V^{N_{\text{A}}+N_{\text{B}}}\frac{(2\pi m_{\text{A}})^{3N_{\text{A}}/2}(2\pi m_{\text{B}})^{3N_{\text{B}}/2}}{\Gamma[3(N_{\text{A}}+N_{\text{B}})/2]} E^{3(N_{\text{A}}+N_{\text{B}})/2-1}
 \end{equation}

The next step consists in generalizing this formalism to the adiabatic grand-isobaric ensemble for the A-B mixture $(\mu_{\text{A}}, \mu_{\text{B}}, P, R)$, {\it i.e.}, with a kinetic energy $K$ defined as $K=R-PV+\mu_{\text{A}} N_{\text{A}} +\mu_{\text{B}} N_{\text{B}} -U(\mathbf{q})$, to obtain the equation analog to Eq.~\ref{genmuPR} for a binary mixture A-B. This yields the general acceptance rule for a MC move from an old configuration $o$ defined by the set $(\mathbf{q},N_{\text{A}}, N_{\text{B}}, V)$ to a new configuration $n$ with the set $(\mathbf{q}',N_{\text{A}}', N_{\text{B}}', V')$ as
\begin{equation}
\begin{array}{lll}
acc(o \rightarrow n) & = & \min \left[ 1, \frac{\rho(\mathbf{q}',N_{\text{A}}', N_{\text{B}}', V')}{\rho(\mathbf{q},N_{\text{A}}, N_{\text{B}}, V)}\right] \\
& = & \min\left[ {1, \frac{ (b_AV')^{N_{\text{A}}'}(b_BV')^{N_{\text{B}}'} N_{\text{A}}! N_{\text{B}}! \Gamma(3(N_{\text{A}}+N_{\text{B}})/2) K_n^{3(N_{\text{A}}'+N_\textbf{B}')/2-1}}{(b_AV)^{N_A}(b_BV)^{N_{\text{B}}} N_{\text{A}}'! N_{\text{B}}'! \Gamma(3(N_{\text{A}}'+N_{\text{B}}')/2) K_o^{3(N_{\text{A}}+N_{\text{B}})/2-1}}} \right]\\
& = & \min\left[ {1, \frac{ (b_AV')^{N_A'}(b_BV')^{N_{\text{B}}'} N_{\text{A}}! N_{\text{B}}! \Gamma(3(N_{\text{A}}+N_{\text{B}})/2) (R-PV'+\mu_{\text{A}} N_{\text{A}}'+\mu_{\text{B}} N_{\text{B}}'-U(\mathbf{q}'))^{3N'/2-1}}{(b_AV)^{N_{\text{A}}}(b_BV)^{N_{\text{B}}} N_{\text{A}}'! N_{\text{B}}'! \Gamma(3(N_{\text{A}}'+N_{\text{B}}')/2) (R-PV+\mu_{\text{A}} N_{\text{A}}+\mu_{\text{B}} N_{\text{B}}-U(\mathbf{q}))^{3N/2-1}}} \right]\\
\end{array}
\label{bimuPR}
\end{equation}
in which $b_A=(2\pi m_{\text{A}} / h^2)^{3/2}$ and $b_B=(2\pi m_{\text{B}}/ h^2)^{3/2}$. 
 
From a practical standpoint, we carry out 4 different types of MC moves corresponding to (i) the translation of a randomly chosen particle (either of type A or B), (ii) the insertion of a particle of type A or B, (iii) the deletion of a particle of type A or B and (iv) a volume change of the system. We provide below explicitly the acceptance rules for each type of move for particles of type A only for conciseness in the case of MC moves of types (i)-(iii). 

The acceptance rule for the translation of a randomly chosen particle of type A from an old (o) configuration to a new (n) configuration is given by 
\begin{equation}
acc(o \rightarrow n)=\min\left[ {1, \frac{(R-PV+\mu_{\text{A}}N_{\text{A}}+\mu_{\text{B}}N_{\text{B}}-U(\mathbf{q}'))^{3(N_{\text{A}}+N_{\text{B}})/2-1}}{{(R-PV+\mu_{\text{A}}N_{\text{A}}+\mu_{\text{B}}N_{\text{B}}-U(\mathbf{q}))^{3(N_{\text{A}}+N_{\text{B}})/2-1}}}} \right]
\end{equation}

The acceptance rule for the insertion of a particle of type A at a random position in the system is given by
\begin{multline}
acc(o \rightarrow n)= \min\left[ 1,\frac{b_A V\Gamma(3(N_{\text{A}}+N_{\text{B}})/2)}{(N_{\text{A}}+1)\Gamma(3(N_{\text{A}}+N_{\text{B}}+1)/2)} \right.\\ \left. \times \frac{[R-PV+\mu_{\text{A}} (N_{\text{A}}+1)+\mu_{\text{B}} N_{\text{B}} -U(\mathbf{q}')]^{3(N_{\text{A}}+N_{\text{B}}+1)/2-1}}{[R-PV+\mu_{\text{A}} N_{\text{A}}+ \mu_{\text{B}} N_{\text{B}} -U(\mathbf{q})]^{3(N_{\text{A}}+N_{\text{B}})/2-1}} \right]
\end{multline}

Similarly, the acceptance rule for the deletion of a particle randomly chosen among the $N_A$ particles, can be written as
\begin{multline}
acc(o \rightarrow n)= \min \left[ 1,\frac{N_A\Gamma(3(N_A+N_B)/2)}{b_A V \Gamma(3(N_A+N_B-1)/2)}  \right. \\ \left. \times  \frac{[R-PV+\mu_A (N_A-1)+ \mu_B N_B -U(\mathbf{q}')]^{3(N_A+N_B-1)/2-1}}{[R-PV+\mu_A N_A+\mu_B N_B -U(\mathbf{q})]^{3(N_A+N_B)/2-1}} \right]
\end{multline}

The acceptance rule for a random volume change of the system is given by
\begin{equation}
acc(o \rightarrow n)=min\left[ {1,\frac{V'^{(N_A+N_B)}[R-PV'+\mu_A N_A + \mu_B N_B -U(\mathbf{q}')]^{3(N_A+N_B)/2-1}}{V^{(N_A+N_B)}[R-PV+\mu_A N_A+\mu_B N_B -U(\mathbf{q})]^{3(N_A+N_B)/2-1}}} \right]
\end{equation}

\subsection{Models}
We use the Lennard-Jones potential to model the interactions between Argon atoms using the following expression
\begin{equation}
\phi(r_{ij})= 4\epsilon \left[ \left( \frac{\sigma}{r_{ij}} \right)^{12}- \left( \frac{\sigma}{r_{ij}} \right)^{6} \right]    
\end{equation}

where $r_{ij}$ is the distance between atom $i$ and atom $j$, $\epsilon$ and $\sigma$ the parameters representing the negative well of depth and the distance for which the potential is equal to zero, respectively. Here, we choose the following set of parameters for our simulations $(\epsilon/k_B)=115.17$~K and $\sigma=3.38$~\AA~\cite{PartIII}. We also use long-range corrections beyond a cutoff at a distance $3\sigma$.~\cite{Allen}. We also use a Lennard-Jones potential to carry out simulations for systems composed of Neon atoms. More specifically, we use the following parameters $(\epsilon/k_B)=33.89$~K and $\sigma=2.79$~\AA~\cite{PartIII}. As previously for Ar, we use the same cutoff and use tail corrections beyond this distance. When we study Ar-Ne mixtures, we use the Lorentz-Berthelot rules~\cite{maitland1981intermolecular,PartIII} to determine the unlike interactions parameters. It gives the following parameters: $\epsilon_{Ar-Ne}=\sqrt{\epsilon_{Ar-Ar}\epsilon_{Ne-Ne}}$ and $\sigma_{Ar-Ne}=\frac{\sigma_{Ar-Ar}+\sigma_{Ne-Ne}}{2}$ with a cutoff at a distance of $r_{cut}=8.37$~\AA~ beyond which tail corrections are applied. \\
As for the metals studied here, both copper and silver are modeled with an embedded-atom (EAM) potential known as the quantum-corrected Sutton-Chen embedded atom model (qSC-EAM)~\cite{finnis1984simple,sutton1990long, mei1991analytic, WAG}. The qSC-EAM potential is composed of two terms: a two-body term and a many-body term: 

\begin{equation}
U= \frac{1}{2} \sum_{i=1}^N \sum_{j \ne i} \epsilon \left( \frac{a}{r_{ij}} \right )^n - \epsilon C \sum_{i=1}^N \sqrt{\rho_i}   
\end{equation}

where $r_{ij}$ is the distance between two atoms $i$ and $j$ and the density term $\rho_i$ is given by

\begin{equation}
\rho_i=\sum_{j \ne i} \left( \frac{a}{r_{ij}}\right)^m    
\end{equation}

We use the parameters obtained by Luo {\it et al.}~\cite{WAG} for Cu, with $\epsilon_{Cu}=0.57921 \times 10^{-2}$~eV, $C_{Cu}=84.843$, $a_{Cu}=3.603$~\AA, $n_{Cu}=10$, and $m_{Cu}=5$ and for Ag, with $\epsilon_{Ag}=0.3945 \times 10^{-2}$~eV, $C_{Ag}=96.524$, $a_{Ag}=4.0691$~\AA, $n_{Ag}=11$, and $m_{Ag}=6$. For each metal, the cutoff distance is set to twice the lattice parameter as in previous work~\cite{desgranges2020central}. When looking at the Cu-Ag mixtures, we use the following rules to determine the interactions between Cu and Ag atoms~\cite{kart2005thermodynamical,kart2004liquid,desgranges2016effect,desgranges2018unusual,desgranges2019can}. $\epsilon_{Cu-Ag}=\sqrt{\epsilon_{Cu-Cu} \epsilon_{Ag-Ag}}$, $m_{Cu-Ag}=\frac{m_{Cu}+m_{Ag}}{2}$, $n_{Cu-Ag}=\frac{n_{Cu}+n_{Ag}}{2}$ and $a_{Cu-Ag}=\frac{a_{Cu}+a_{Ag}}{2}$. 

\subsection{Simulation details}

To determine the properties of mixing, we carry out two types of simulations in the grand-isobaric adiabatic ensemble for single-component systems, and for two-component systems. For single-component systems, we use $(\mu,P,R)$ simulations, in which $\mu$ is the chemical potential, $P$ the pressure and $R$ the Ray or heat function. The heat function provides access to the entropy of the system through the relation $R=TS$. For an A-B binary mixture, we carry out simulations in the $(\mu_A,\mu_B,P,R)$ ensemble. Here, $\mu_A$ and $\mu_B$ denote the chemical potentials for components A and component B, respectively. Since we implement simulations in the grand-isobaric adiabatic ensemble within a Monte Carlo (MC) framework, we perform the following types of MC moves with the attempt probabilities as follows: (i) 33\% of the attempted moves are random displacements of an atom, (ii) 33\% are insertions of atoms at random locations within the system, (iii) 33\% are deletions of randomly selected atoms, and (iv) 1\% are random volume changes for the entire system. For each set of conditions, we carry out two successive runs. We first perform a run of $10^8$ MC steps to allow the system to relax and the simulation to converge toward equilibrium. We then carry out a production run of $10^8$ MC steps over which averages are calculated. Statistical uncertainties are evaluated using the standard block averaging technique over blocks of $5 \times 10^7$ MC steps. Finally, the average temperature of the system, $<T>$, is evaluated through the equipartition principle. We use the following expression for the kinetic energy $K$ in a two-component system, $K=R-PV+\mu_A N_A +\mu_B N_B -U$ and calculate the average temperature as $<T>= \frac{2<K>}{3k_B<N_A+N_B>}$. This, in turn, allows for the determination of the entropy of the system through the equation $<S>=R/<T>$. Throughout the paper, we note as $\bar{Y}$ the molar property for any extensive quantity $Y$. To test the accuracy and reliability of the grand-isobaric adiabatic ensemble approach, we also carry out MC simulations in the $(N,P,T)$ ensemble for single-component systems, as well as in the $(N_A,N_B,P,T)$ ensemble for binary mixtures. For single-component systems, we work with $N=500$~atoms, and use the following probabilities for the various types of MC moves: (i) 99\% of attempted moves are translations of a single, randomly chosen, atom, and (ii) 1\% of attempted moves are random volume changes. For two-component systems, we use a total number of atoms of $N_A+N_B=500$ in the simulations, and carry out MC moves with the same probabilities as for single-component systems. 

\section{Results and discussion}

\subsection{The Argon-Neon system}

\subsubsection{Single-component systems}
We first present results for the single-component systems, Ar and Ne. We perform 8 different ($\mu$, P,R) simulations for which $P$ and $R$ are held constant. More specifically, in the case of Ar, we vary $\mu$ between $\mu=-230$~kJ/kg to $\mu=-550$~kJ/kg along the isobar $P=445$~bar for a value of the heat function set to $R/k_B=8 \times 10^5$~K. For each value of $\mu$, we report the corresponding number of atoms $<N>$, temperature $<T>$, specific volume $\bar{V}=\frac{<V>}{<N>}$, enthalpy $\bar{H}=\frac{<H>}{<N>}$, Ray energy $\bar{R}=\frac{R}{<N>}$ and entropy $\bar{S}=\frac{<S>}{<N>}$. We present our results in Table~\ref{Tab1}. 

\begin{table}[ht]
\centering
\begin{tabular}{cccccccc}
\hline\hline
$\mu$~(kJ/kg)~&$<N>$~&$<T>$~(K)~&$\bar{V}$~(cm$^3$/g)~&$\bar{H}$~(kJ/kg)~&$\bar{R}$~(kJ/kg)~&$\bar{S}$~(kJ/kg/K)~&$\bar{S}_{id}$~(kJ/kg/K)\\
\hline
-230 & 1067.3 & 106.1 & 0.702 & -74.11 & 155.9 & 1.469 &  2.139\\
-250 & 880.1 & 119.2 & 0.734 & -60.94 & 189.1 & 1.586 &  2.186\\
-300 & 870.4 & 148.8 & 0.816 & -31.84 & 268.2 & 1.803 &  2.277\\
-350 & 754.2 & 175.2 & 0.908 & -5.69 & 344.3 & 1.968 & 2.349\\
-400 & 690.7 & 199.8 & 1.011 & 18.52 & 418.5 & 2.096 & 2,414\\
-450 & 654.4 & 223.1 & 1.124 & 41.06 & 491.0 & 2.203 & 2.470\\
-500 & 632.8 & 245.3 & 1.244 & 62.02 & 562.0 & 2.296 & 2.520\\
-550 & 618.9 & 266.8 & 1.367 & 81.39 & 631.4 & 2.366 & 2.567\\
\hline\hline
\end{tabular}
\caption{Argon: $(\mu,P,R)$ simulation results along the $P=445$~bar isobar and for $R/k_B=8 \times 10^5$~K. $\bar{S}_{id}$ indicates the value for the ideal gas entropy provided by the Sackur=Tetrode equation.}
\label{Tab1}
\end{table}

Table~\ref{Tab1} shows the results obtained for Argon. For instance, we find that for $\mu=-250$~kJ/kg, $<T>=119.2$~K, $\bar{V}=0.734$~cm$^3$/g, $\bar{H}=-60.94$~kJ/kg, $\bar{R}=189.1$~kJ/kg and $\bar{S}=1.586$~kJ/kg/K. This is in excellent agreement with the experimental data~\cite{Vargaftik}, with, for $T=120$~K and $P=445$~bar, a specific volume of $0.741$~cm$^3$/g, an enthalpy of $-61.2$~kJ/kg, and an entropy of $1.585$~kJ/kg/K. Table~\ref{Tab1} shows that, as the chemical potential decreases, the number of Ar atoms decreases, the temperature increases and the specific volume increases, meaning that the system becomes less and less dense. This is in line with the increase in enthalpy and entropy, that result from the decreased number of interactions between Ar atoms and the loss of organization in the system. Table~\ref{Tab1} also provides the value taken by the ideal gas entropy according to the Sackur-Tetrode equation
\begin{equation}
\bar{S}_{id}= k_B \ln \left[ \left( 2 \pi m k_B T \over h^2 \right)^{3/2} {V e^{5/2} \over N}  \right]    
\end{equation}
The ideal gas entropy $\bar{S}_{id}$ is found to be larger than the molar entropy $\bar{S}$. This results from the attractive interactions that take place in liquid Argon. As shown in Table~\ref{Tab1}, this effect decreases as the density decreases or, equivalently, the specific volume increases, leading to the narrowing of the gap between $\bar{S}_{id}$ and $\bar{S}$.

\begin{figure}
\begin{center}
\includegraphics*[width=12cm]{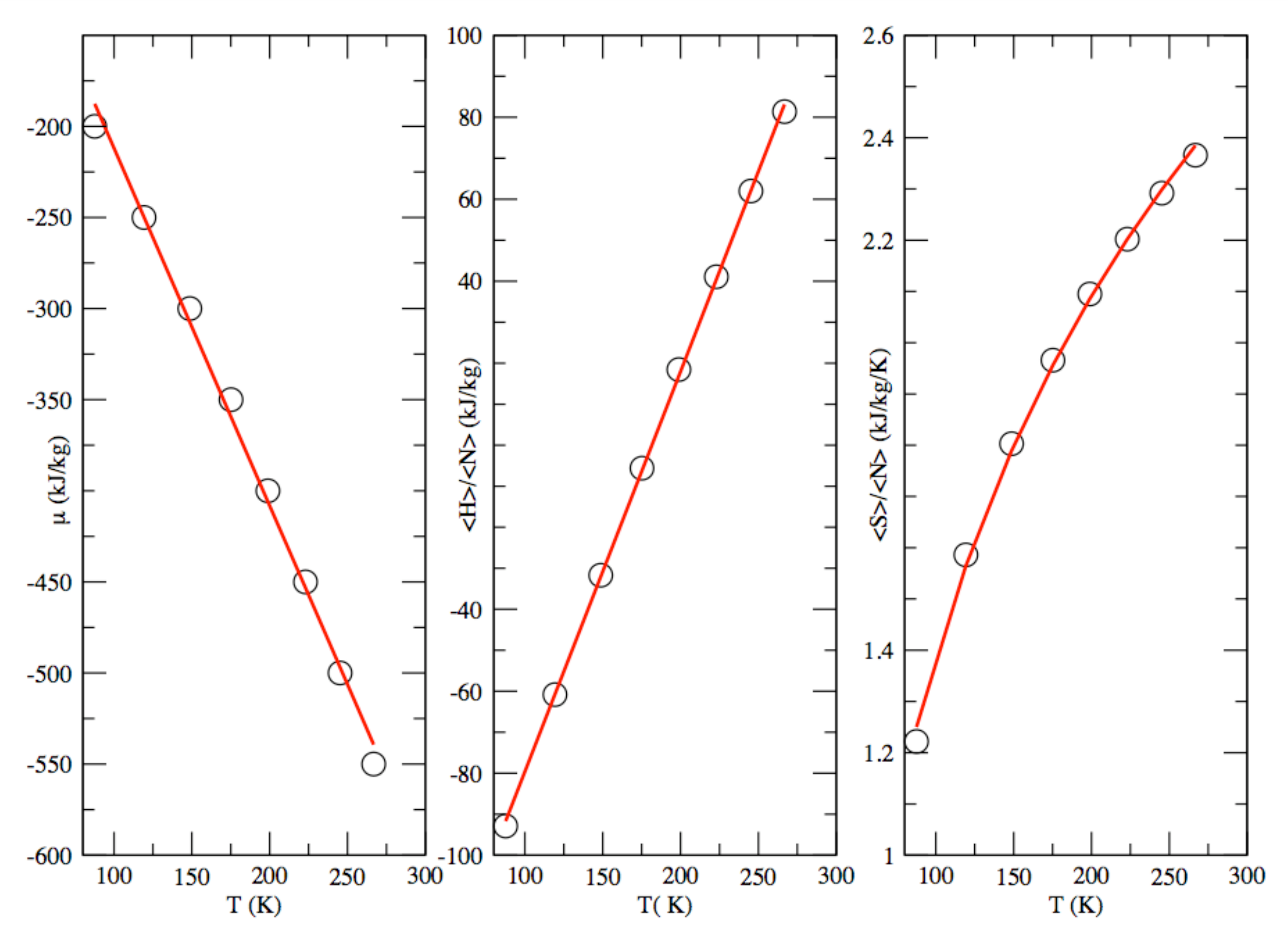}
\end{center}
\caption{Argon along the $P=445$~bar isobar: (a) Chemical potential $\mu$ as a function of the temperature $T$, with a linear fit shown as a red line, (b) Enthalpy $\bar{H}=\frac{<H>}{<N>}$ as a function of $T$, with a linear fit to the simulation results shown as a red line, (c) Entropy $\bar{S}=\frac{<S>}{<N>}$ against $<T>$, with a logarithmic fit to the simulation results shown in red.}
\label{Fig1}
\end{figure}

We show in Fig~\ref{Fig1} the variation of the chemical potential, enthalpy, and entropy as a function of $T$ along the $P=445$~bar isobar. We also provide in Fig~\ref{Fig1} a linear fit for $\mu(T)$, $\mu(T)$~(kJ/kg)$=-15.692 - 1.963 \times T$. Given the thermodynamic relation $\mu=\bar{H}-T\bar{S}$, this gives an estimate for the average entropy over this temperature interval of $1.963$~kJ/kg/K in reasonable agreement with the range of experimental $\bar{S}$ values of $1.4-2.4$~kJ/kg/K for this temperature interval~\cite{Vargaftik}. We also show in Fig~\ref{Fig1} a linear fit for $\bar{H}(T)$, with $\bar{H}(T)$~(kJ/kg)$=-177.13 + 0.976 \times T$, as well as a logarithmic fit for $\bar{S}(T)$: $\bar{S}(T)$~(kJ/kg)$=-3.307 + 1.019 \ln (T)$. To further assess the accuracy of the fits, we run a separate ($\mu,P,R$) simulation for the following set: $\mu=-233$~kJ/kg, $P=445$~bar and $R/k_B=8 \times 10^5$~K. We find $T=108.2$~K, $\bar{H}=-72.06$~kJ/kg, and $\bar{S}=1.488$~kJ/kg/K. These results are in excellent agreement with the values found from the fits developed above, {\it i.e.}, $T=110.78$~K, $\bar{H}=-69.05$~kJ/kg, and $\bar{S}=1.489$~kJ/kg/K. This confirms that the fits given above capture the variation of thermodynamic properties in Argon for the thermodynamic parameters studied in this work.

\begin{table}[ht]
\centering
\begin{tabular}{cccccccc}
\hline\hline
$\mu$~(kJ/kg)~&$<N>$~&$<T>$~(K)~&$\bar{V}$~(cm$^3$/g)~&~$\bar{H}$~(kJ/kg)~&$\bar{R}$~(kJ/kg)~&$\bar{S}$~(kJ/kg/K)~&$\bar{S}_{id}$~(kJ/kg/K)\\
\hline
-250 & 987.1 & 100.0 & 1.355 & 83.60 & 333.6 & 3.335 & 3.767\\
-300 & 813.9 & 114.6 & 1.512 & 104.62 & 404.6 & 3.530 & 3.896\\
-350 & 695.1 & 128.3 & 1.663 & 123.73 & 473.7 & 3.692 & 4.005\\
-400 & 608.1 & 141.7 & 1.806 & 141.49 & 541.5 & 3.821 & 4.100\\
-450 & 541.7 & 154.5 & 1.941 & 157.92 & 607.9 & 3.934 & 4.184\\
-500 & 488.7 & 167.1 & 2.076 & 173.86 & 673.9 & 4.032 & 4.260\\
-550 & 445.7 & 179.5 & 2.202 & 188.84 & 738.8 & 4.116 & 4.328\\
-600 & 409.9 & 191.3 & 2.328 & 203.27 & 803.3 & 4.199 & 4.390\\
\hline\hline
\end{tabular}
\caption{Neon: $(\mu,P,R)$ simulation results along the $P=445$~bar isobar and for $R/k_B=8 \times 10^5$~K.}
\label{Tab2}
\end{table}

Next, we turn to the second single-component system, Neon. We follow the same protocol as for Argon and present in Table~\ref{Tab2} the results obtained from ($\mu,P,R$) simulations. We observe the same general behavior as for Argon. As $\mu$ decreases, the number of Ne atoms decreases, $T$ increases and the specific volume increases. The main difference with Ar for the set of thermodynamic conditions studied in this work is that the enthalpy is always positive, which results from the fact that the system is a supercritical fluid under these conditions. As for Argon, the results found with the $(\mu,P,R)$ method are in very good agreement with those found using $(N,P,T)$ simulations. For instance, we find that for $\mu=-450$~kJ/kg, $<T>=154.5$~K, $\bar{V}=1.941$~cm$^3$/g and $\bar{H}=157.92$~kJ/kg. This is in excellent agreement with simulation results we obtain in the $(N,P,T)$ ensemble for $T=154.5$~K and $P=445$~bar, with a specific volume estimated at $1.938$~cm$^3$/g and an enthalpy of $157.97$~kJ/kg. Table~\ref{Tab2} also provides a comparison with the ideal gas entropy obtained from the Sackur-Tetrode equation. Since, under the thermodynamic conditions used in this work, Neon is a supercritical fluid, the interactions play a lesser role than for Argon, which is a liquid under these thermodynamic conditions. As a result, the molar entropy of Neon is found to be close to the ideal gas entropy under these conditions.

\begin{figure}
\begin{center}
\includegraphics*[width=12cm]{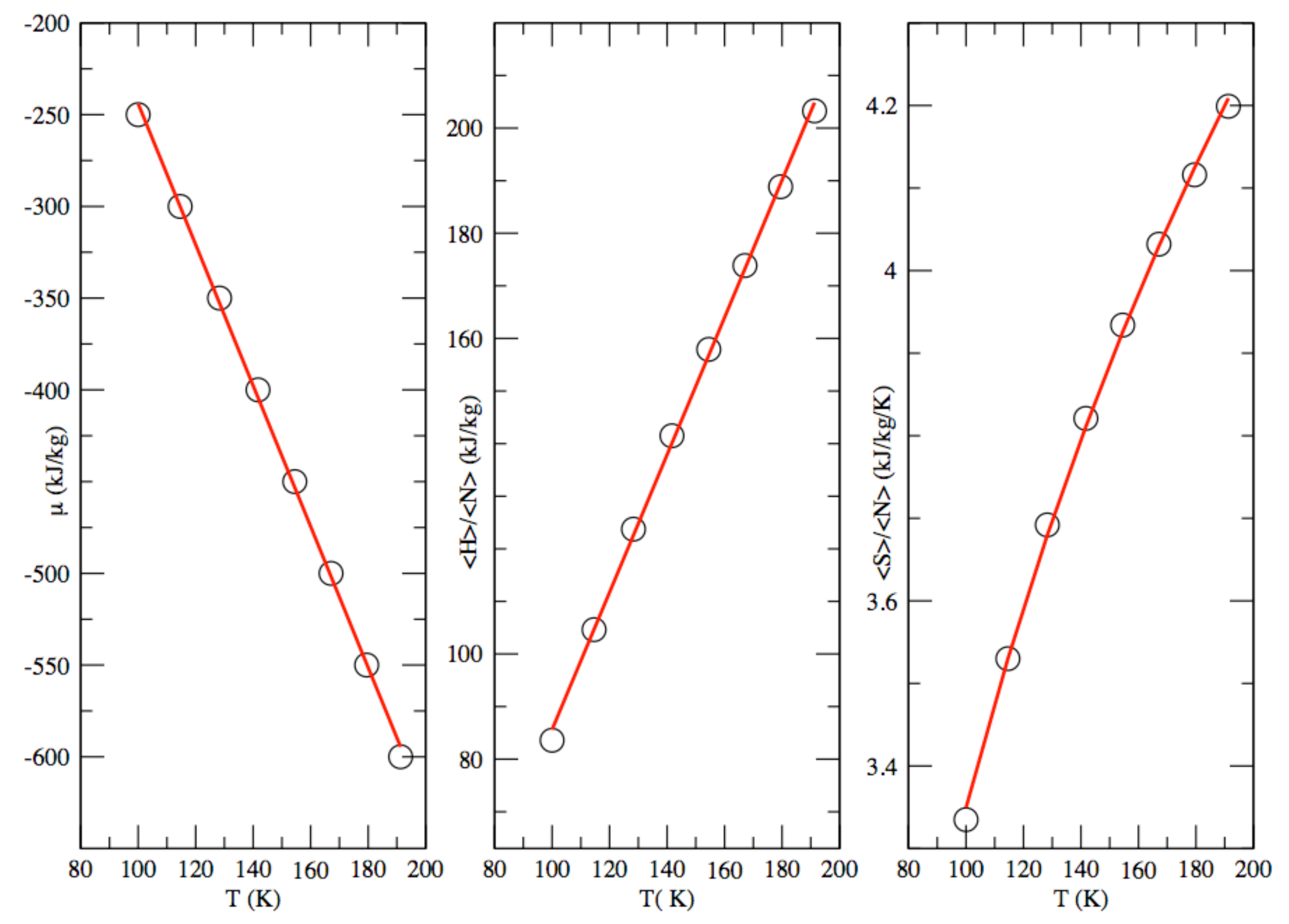}
\end{center}
\caption{Neon along the $P=445$~bar isobar: (a) ($\mu,P,R$) simulation results for $\mu$ against $T$, with a linear fit shown in red, (b) Enthalpy $\bar{H}=\frac{<H>}{<N>}$ against $T$, with a linear fit shown as a red line, and (c) Entropy $\bar{S}=\frac{<S>}{<N>}$ against $<T>$, with a logarithmic fit shown as a red solid line.}
\label{Fig2}
\end{figure}

We also show in Fig~\ref{Fig2} the variation of $\mu$, $\bar{H}$, and $\bar{S}$ as a function of $T$ along the $P=445$~bar isobar. As for Ar, we provide a linear fit for $\mu(T)$: $\mu(T)$~(kJ/kg)$=139.89 - 3.840 \times T$, which gives us an average entropy of $3.840$~kJ/kg/K over the temperature interval. The linear fit for $H(T)$ is given by $\bar{H}(T)$~(kJ/kg)$=-44.95 + 1.306 \times T$. Finally, a logarithmic fit for $S(T)$ gives $\bar{S}(T)$~(kJ/kg/K)$=-2.751 + 1.325 \ln (T)$. To further assess the accuracy of the fits, we perform a separate ($\mu,P,R$) simulation with $\mu=-285$~kJ/kg, $P=445$~bar and $R/k_B=8 \times 10^5$~K. We obtain $<T>=110.6$~K, $\bar{H}=98.78$~kJ/kg and $\bar{S}=3.471$~kJ/kg/K. These results are in good agreement with the values from the above fits, {\it i.e.}, $T=110.6$~K, $\mu=-284.8$~kJ/kg, $\bar{H}=98.54$~kJ/kg and $\bar{S}=3.483$~kJ/kg/K, which confirms the validity of the $(\mu,P,R)$ simulation method.

\subsubsection{Thermodynamic properties for the Ar-Ne mixture}

We now turn to the study of the Ar-Ne mixture, and run simulations in the ($\mu_{Ne}, \mu_{Ar}, P, R$) ensemble. To better understand how simulations in this ensemble work, we start by investigating the role played by $R$, the heat function or Ray energy, when two components are present in the system. To this end, we hold $\mu_{Ne}$, $\mu_{Ar}$ and $P$ constant and gradually vary $R$. Results are given in Table~\ref{Tab3}. In this Table, the mole fraction in Ne is calculated as $<x_{Ne}>=\frac{<N_{Ne}>}{<N_{Ne}>+<N_{Ar}>}$, the temperature as $<T>=\frac{2<K>}{3<N_{tot}>k_B}$, with $N_{tot}=N_{Ne}+N_{Ar}$, and the density as $<\rho>=\frac{m_{Ne} <N_{Ne}>+m_{Ar} <N_{Ar}>}{<V>}$. Datasets available for comparison include the measurements by Streett~\cite{streett1967liquid}, as well as the reference model developed by Tkaczuk {\it et al.} \cite{tkaczuk2020equations}.

\begin{table}[ht]
\centering
\begin{tabular}{cccccccc}
\hline\hline
$\frac{R}{k_B}$~&~$<N_{Ne}>$~&~$<N_{Ar}>$~&~$<x_{Ne}>$~&~$<T>$~&$<\rho>$~&$\bar{H}$~&~$\bar{S}$\\
(K) & - & - & - & K & (g/cm$^3$) & (kJ/kg) & (kJ/kg/K) \\
\hline
$2 \times 10^5$~& 106.5 & 106.0 & 0.501 & 110.8 & 1.128 & -11.6 & 4.697 \\
$3 \times 10^5$~& 159.7 & 159.1 & 0.501 & 110.9 & 1.129 & -11.8 & 4.694 \\
$4 \times 10^5$~& 212.2 & 212.9 & 0.499 & 110.9 & 1.130 & -12.3 & 4.687 \\
$5 \times 10^5$~& 266.4 & 264.7 & 0.502 & 110.8 & 1.128 & -11.5 & 4.701 \\
$6 \times 10^5$~& 318.6 & 319.0 & 0.500 & 110.8 & 1.130 & -12.2 & 4.693 \\
$7 \times 10^5$~& 372.4 & 371.4 & 0.501 & 110.8 & 1.129 & -11.9 & 4.697 \\
$8 \times 10^5$~& 425.5 & 424.5 & 0.501 & 110.8 & 1.129 & -11.9 & 4.697 \\
$9 \times 10^5$~& 478.9 & 477.4 & 0.500 & 110.9 & 1.129 & -11.9 & 4.690 \\
\hline\hline
\end{tabular}
\caption{Ar-Ne mixture along the $P=445$~bar isobar: $(\mu_{Ne},\mu_{Ar},P,R)$ simulation results for $x_{Ne}=0.5$, {\it i.e.}, for $\mu_{Ne}=-203$~kJ/kg and $\mu_{Ar}=-330$~kJ/kg, for different values of the heat function $R$.}
\label{Tab3}
\end{table}

\begin{figure}
\begin{center}
\includegraphics*[width=12cm]{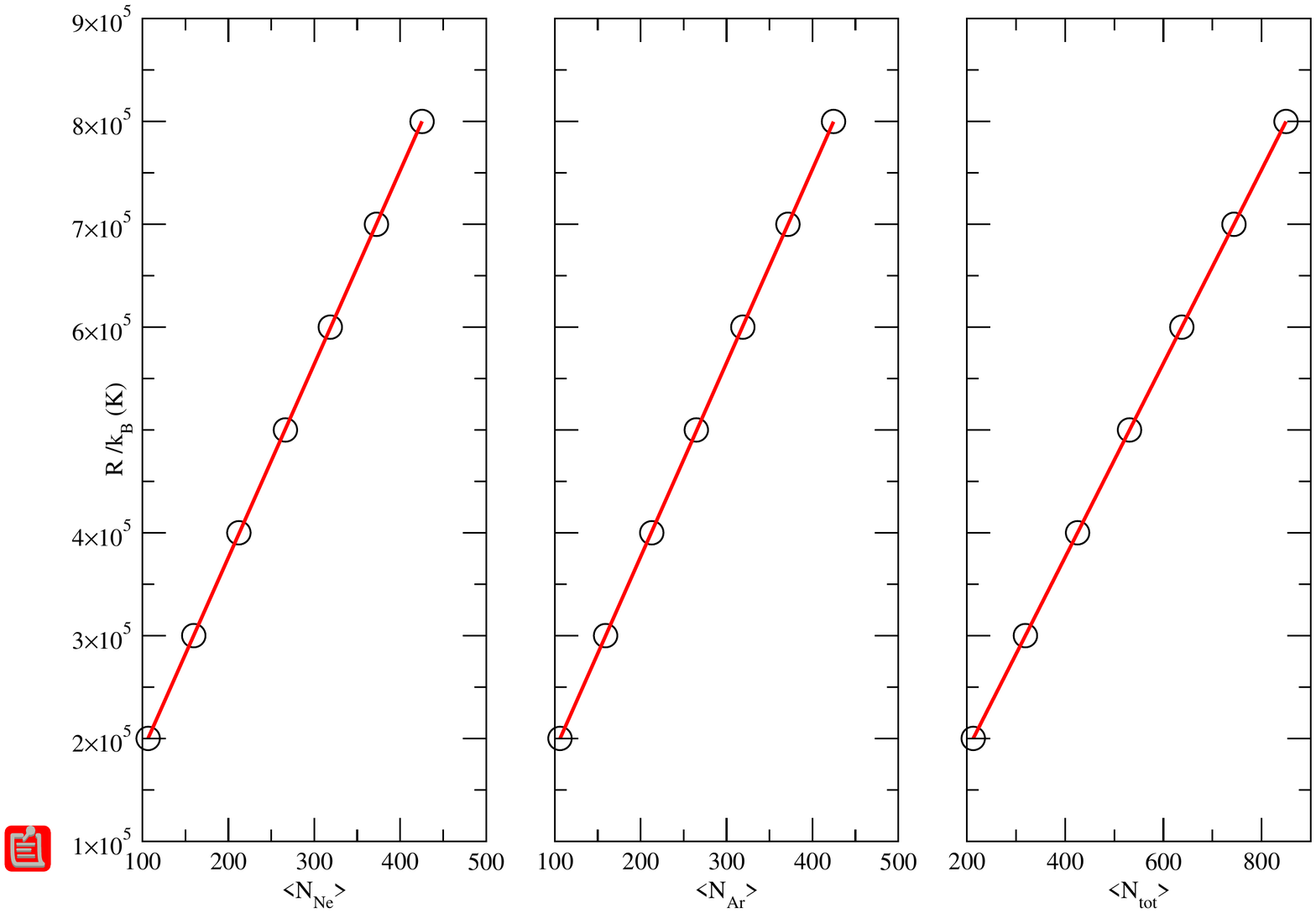}
\end{center}
\caption{Ar-Ne mixture. Ray function $R$ in function of (a) the number of Ne atoms, $<N_{Ne}>$, (b) the number of Ar atoms $<N_{Ar}>$, and (c) the total number of particles $<N_{tot}>$, present in the $x_{Ne}=0.5$ mixture.}
\label{Fig3}
\end{figure}

Results from Table~\ref{Tab3} show that, at fixed $P$ and for a given set of chemical potentials ($\mu_{Ne}$, $\mu_{Ar}$), increasing the value of $R$ leads to an increase in the number of atoms for the two components of the mixture, $<N_{Ne}>$ and $<N_{Ar}>$, and thus in the total number of atoms in the system $<N_{tot}>$. For instance, multiplying by $5$ the value of $R$ leads to a $5$-fold increase in $<N_{Ne}>$ and $<N_{Ar}>$. Interestingly, we observe a linear dependence on $R$ for all numbers of atoms, $<N_{Ne}>$, $<N_{Ar}>$ and $<N_{tot}>$ (see Fig~\ref{Fig3}). This implies that, once the ($\mu_{Ne}, \mu_{Ar}, P, R$) simulations have converged, and regardless of the value set for $R$, the intensive thermodynamic properties all converge towards the same values. For instance, for $R/k_B=2 \times 10^5$~K, the system converges towards a temperature of $110.8$~K a density of $1.128$~g/cm$^3$, an enthalpy of $-11.6$~kJ/kg and an entropy of $4.697$~kJ/kg/K. For $R/k_B=9 \times 10^5$~K, the system reaches at convergence $110.9$~K for $<T>$, $1.129$~g/cm$^3$ for $<\rho>$, $-11.9$~kJ/kg/K for $<\bar{H}>$, and $4.690$~kJ/kg/K for $<\bar{S}>$. The sets of results obtained for the these two $R$ values are within the statistical uncertainty of the simulations, which are of $0.5$~K, $0.008$g/cm$^3$, $0.2$~kJ/kg, and $0.15$~kJ/kg/K for the temperature, density, enthalpy, and entropy, respectively.

In the rest of the paper, we present results obtained from ($\mu_{Ne}, \mu_{Ar},P,R$) simulations at $P=445$~bar and $R/k_B=3 \times 10^5$~K, as they are found from Table~\ref{Tab3} to provide accurate results for a reasonably small total number of atoms in the system. To obtain mixture properties for mole fractions in Ne spanning the entire range from 0.1 to 0.9, we vary both $\mu_{Ne}$ and $\mu_{Ar}$ and select conditions for which the temperature of the system has converged towards $<T>=110.8 \pm 0.5$~K. We present the results in Table~\ref{Tab4}.

\begin{table}[ht]
\centering
\begin{tabular}{ccccccc}
\hline\hline
$<x_{Ne}>$&~$\mu_{Ne}$(kJ/kg)&~$\mu_{Ar}$(kJ/kg)~&$<N_{Ne}>$~&~$<N_{Ar}>$~&~$<T>$~(K)~&~$<\rho>$~(g/cm$^3$) \\
\hline
0.1 & -187 & -251 & 37.2 & 316.5 & 110.5 & 1.361  \\
0.2 & -183 & -267 & 67.9 & 275.3 & 110.3 & 1.320  \\
0.3 & -186 & -285 & 98.9 & 235.9 & 110.3 & 1.268  \\
0.4 & -193 & -307 & 128.8 & 195.9 & 110.8 & 1.202  \\
0.5 & -203 & -330 & 159.7 & 159.1 & 110.9 & 1.129  \\
0.6 & -215 & -356 & 188.2 & 126.1 & 110.9 & 1.049  \\
0.7 & -229 & -386 & 217.5 & 94.5 & 110.7 & 0.963  \\
0.8 & -246 & -424 & 247.8 & 62.3 & 110.9 & 0.866  \\
0.9 & -264 & -478 & 282.1 & 30.6 & 110.6 & 0.770  \\
\hline\hline
\end{tabular}
\caption{Ar-Ne mixture. Results from ($\mu_{Ne}$,$\mu_{Ar}$,P,R) simulations along the isobar $P=445$~bar with $R/k_B= 3 \times 10^5$, and $<T>=110.8 \pm 0.5$~K}
\label{Tab4}
\end{table}

\begin{figure}
\begin{center}
\includegraphics*[width=12cm]{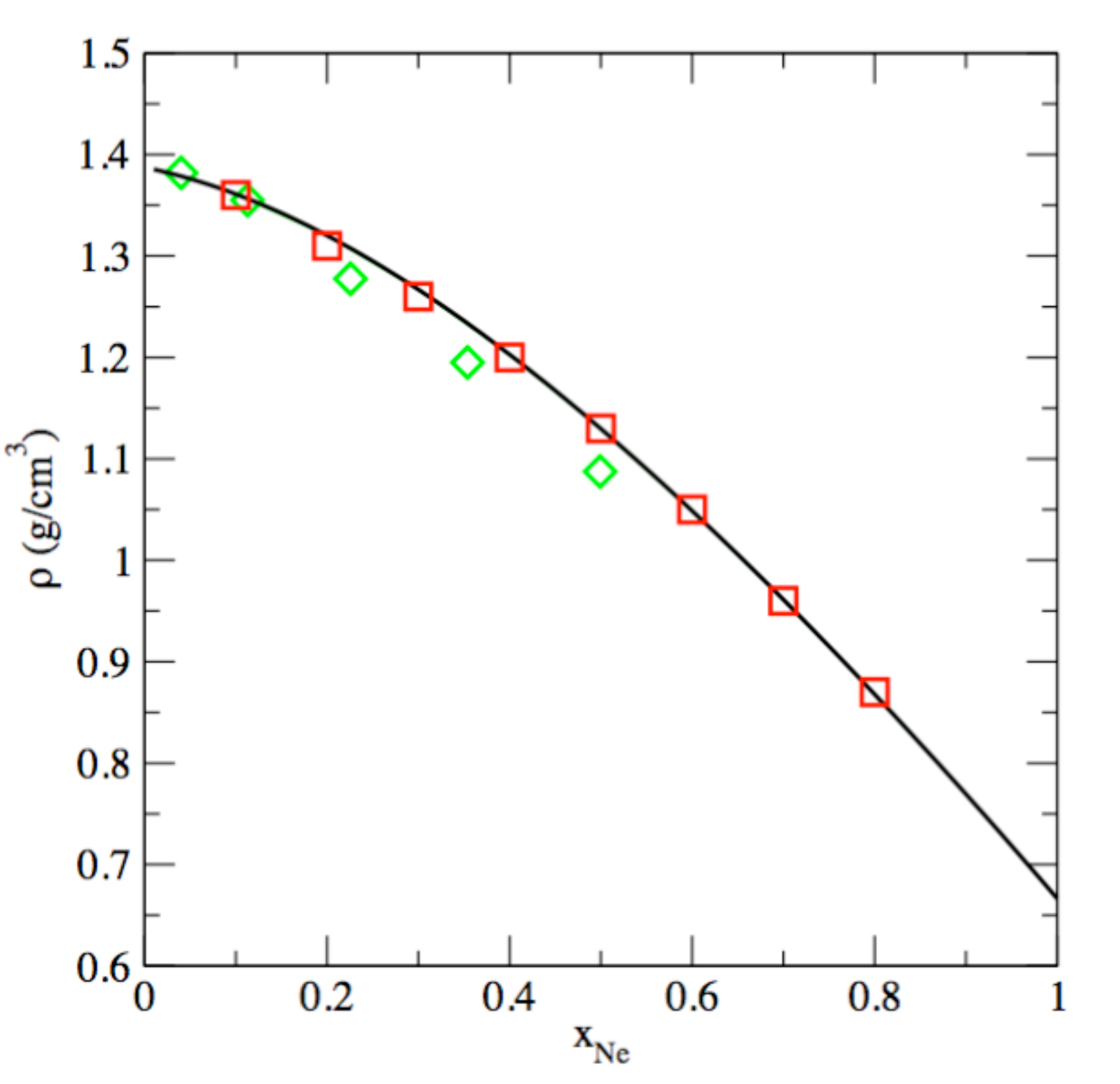}
\end{center}
\caption{Density against the mole fraction in Ne $x_{Ne}$ for the Ar-Ne mixture along the isobar $P=445$~bar. $(N_{Ne}, N_{Ar}, P, T)$ simulation results are shown as squares, while experimental data are shown as diamonds. The experimental data~\cite{streett1967liquid} are given for a temperature $T=110.78$~K and a pressure of $P=6500$~psia or, equivalently, $P=448$~bar.}
\label{Fig4}
\end{figure}

To assess the accuracy of the ($\mu_{Ne}$, $\mu_{Ar}$, P, R) simulations, we carry out a simulation in the ($N_{Ne}$, $N_{Ar}$, P, T) ensemble at $110.8$~K and $P=445$~bar and for a mole fraction in Ne of 0.5. We find an average density of $1.13$~g/cm$^3$ in good agreement with the ($\mu_{Ne}$, $\mu_{Ar}$, P, R) simulation results of $1.129$~g/cm$^3$. This is a first validation of the grand-isobaric adiabatic ensemble for mixtures. By fitting the ($\mu_{Ne}$, $\mu_{Ar}$, P, R) simulation results, we obtain the following equation for the density of the mixture
\begin{equation}
\rho~(g/cm^3)=1.387-0.183x_{Ne}-0.838x^2_{Ne}+0.394x^3_{Ne}-0.093x^4_{Ne}
\label{fitrhomNeAr}
\end{equation}
and test the fit against with the available experimental data under these conditions~\cite{streett1967liquid} and the results obtained from $(N_{Ne}, N_{Ar}, P, T)$ simulations. As shown in in Fig.~\ref{Fig4}, there is a good agreement between the three sets of data over the entire range of compositions, which shows that Eq.~\ref{fitrhomNeAr} provides an accurate model for the density of the Ne-Ar mixture.

\begin{figure}
\begin{center}
\includegraphics*[width=12cm]{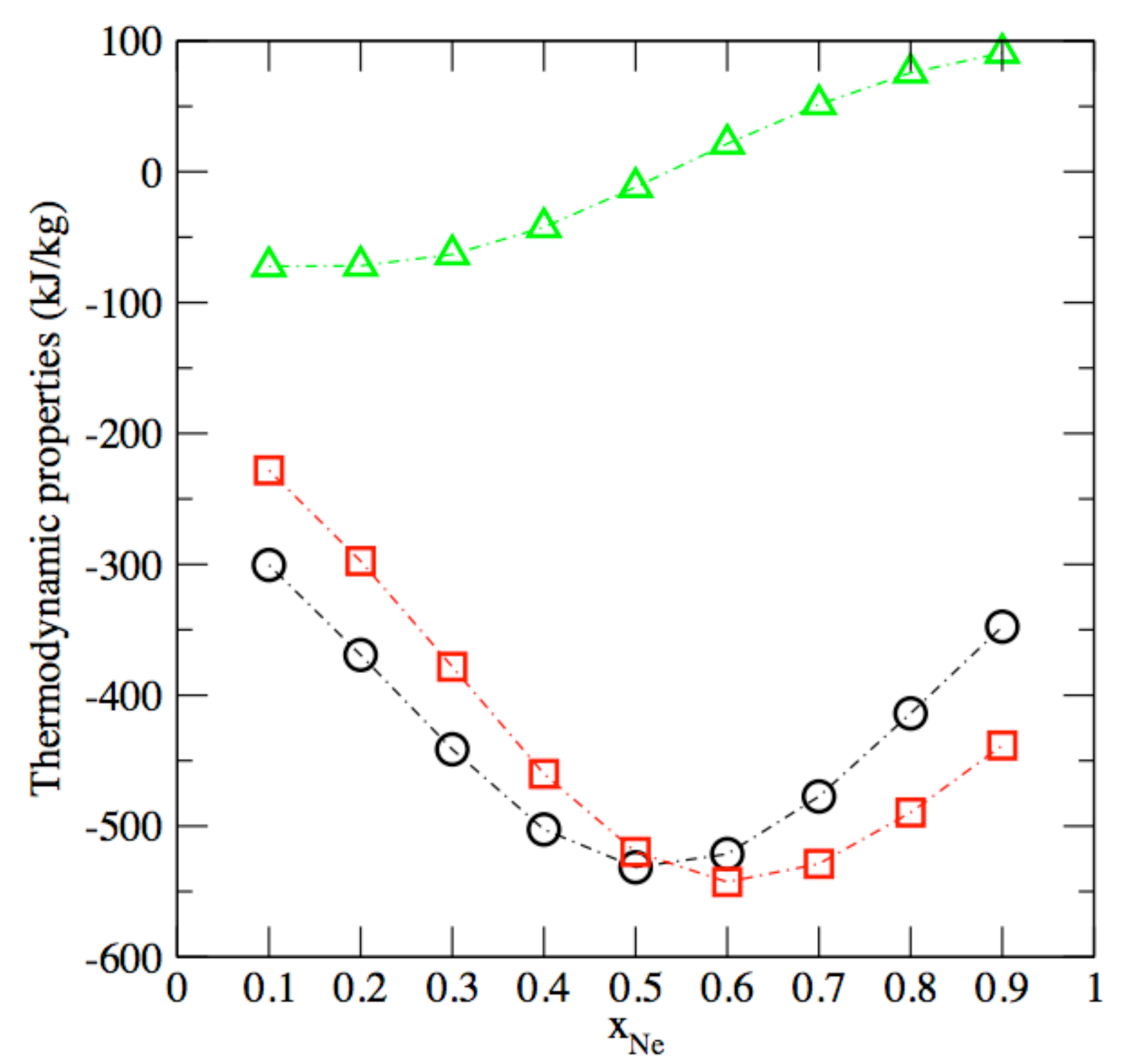}
\end{center}
\caption{Ne-Ar mixture at $110.8$~K and $445$~bar. $(\mu_{Ne}, \mu_{Ar}, P, R)$ simulation results for the enthalpy (green triangles), product $-T\bar{S}$ (red circles) and Gibbs free energy (black circles) against $x_{Ne}$.}
\label{Fig5}
\end{figure}

We now examine the variations of the thermodynamic properties of the mixture as a function of the mole fraction in Ne. We focus in Fig.~\ref{Fig5} on the plots for enthalpy $\bar{H}$, product $-T\bar{S}$ and Gibbs free energy, calculated as $\bar{G}=\bar{H}-T\bar{S}$. For a low $x_{Ne}$, the mixture exhibits the signature of a liquid with a negative enthalpy. Then, as $x_{Ne}$ increases, enthalpy increases since Neon is a supercritical fluid under these conditions and is associated with a positive enthalpy. On the other hand, we observe a non-monotonic behavior, with a maximum for $\bar{S}$ and thus a minimum for $-T\bar{S}$ and for the Gibbs free energy $\bar{G}$. We find that the entropy reaches a maximum, and the Gibbs free energy a minimum, close to an equimolar fraction for the two components of the mixture.

We now move on to the determination of the thermodynamic properties of mixing. For any thermodynamic property $\bar{Y}$, we evaluate the property of mixing $\Delta \bar{Y}_{mix}$ from the property determined for the mixture $\bar{Y}_{m}$, and the properties for the single-component system $\bar{Y}_{Ne}$ and $\bar{Y}_{Ar}$, determined under the same conditions of $T$ and $P$. Specifically, we obtain here
\begin{equation}
\begin{tabular}{c}
$\Delta \bar{H}_{mix}=\bar{H}_{m} - x_{Ne} \bar{H}_{Ne} - x_{Ar} \bar{H}_{Ar}$\\
$\Delta \bar{S}_{mix}=\bar{S}_{m} - x_{Ne} \bar{S}_{Ne} - x_{Ar} \bar{S}_{Ar}$\\
$\Delta \bar{G}_{mix}=\bar{G}_{m} - x_{Ne} \mu_{Ne} - x_{Ar} \mu_{Ar}$\\
\end{tabular}
\label{Mixprop}
\end{equation}

\begin{table}[ht]
\centering
\begin{tabular}{cccccccc}
\hline\hline
$<x_{Ne}>$~&$\bar{H}_{m}$~&$\Delta \bar{H}_{mix}$~&$\bar{S}_{m}$~&$\Delta \bar{S}_{mix}$~&$\bar{G}_{m}$~&$\Delta \bar{G}_{mix}$\\
- & (kJ/kg) & (kJ/kg) & (kJ/kg/K) & (kJ/kg/K) & (kJ/kg) & (kJ/kg) \\ 
\hline
0.1 & -72.33 & -12.20 & 2.060 & 0.465 & -300.51 & -55.55 \\
0.2 & -71.83 & -21.60 & 2.685 & 0.974 & -369.26 & -118.72\\
0.3 & -63.19 & -24.02 & 3.414 & 1.572 & -441.38 & -185.91 \\
0.4 & -42.48 & -15.71 & 4.153 & 2.165 & -502.57 & -241.20 \\
0.5 & -11.79 & 0.92 & 4.691 & 2.537 & -531.47 & -265.35 \\
0.6 & 21.34 & 18.03 & 4.899 & 2.555 & -521.35 & -250.22 \\
0.7 & 51.39 & 29.63 & 4.773 & 2.212 & -477.40 & -201.60 \\
0.8 & 75.67 & 32.45 & 4.420 & 1.605 & -413.97 & -132.55 \\
0.9 & 90.78 & 22.26 & 3.957 & 0.844 & -347.61 & -62.11 \\
\hline\hline
\end{tabular}
\caption{Ar-Ne mixture at $T=110.8$~K and $P=445$~bar. $(\mu_{Ne}, \mu_{Ar}, P. R)$ results for the thermodynamic properties of the mixture $\bar{Y}_m$ and the thermodynamic properties of mixing $\Delta \bar{Y}_{mix}$. Properties are given in kJ/kg for enthalpy and Gibbs free energy, and in kJ/kg/K for entropy.}
\label{Tab5}
\end{table}

\begin{figure}
\begin{center}
\includegraphics*[width=10cm]{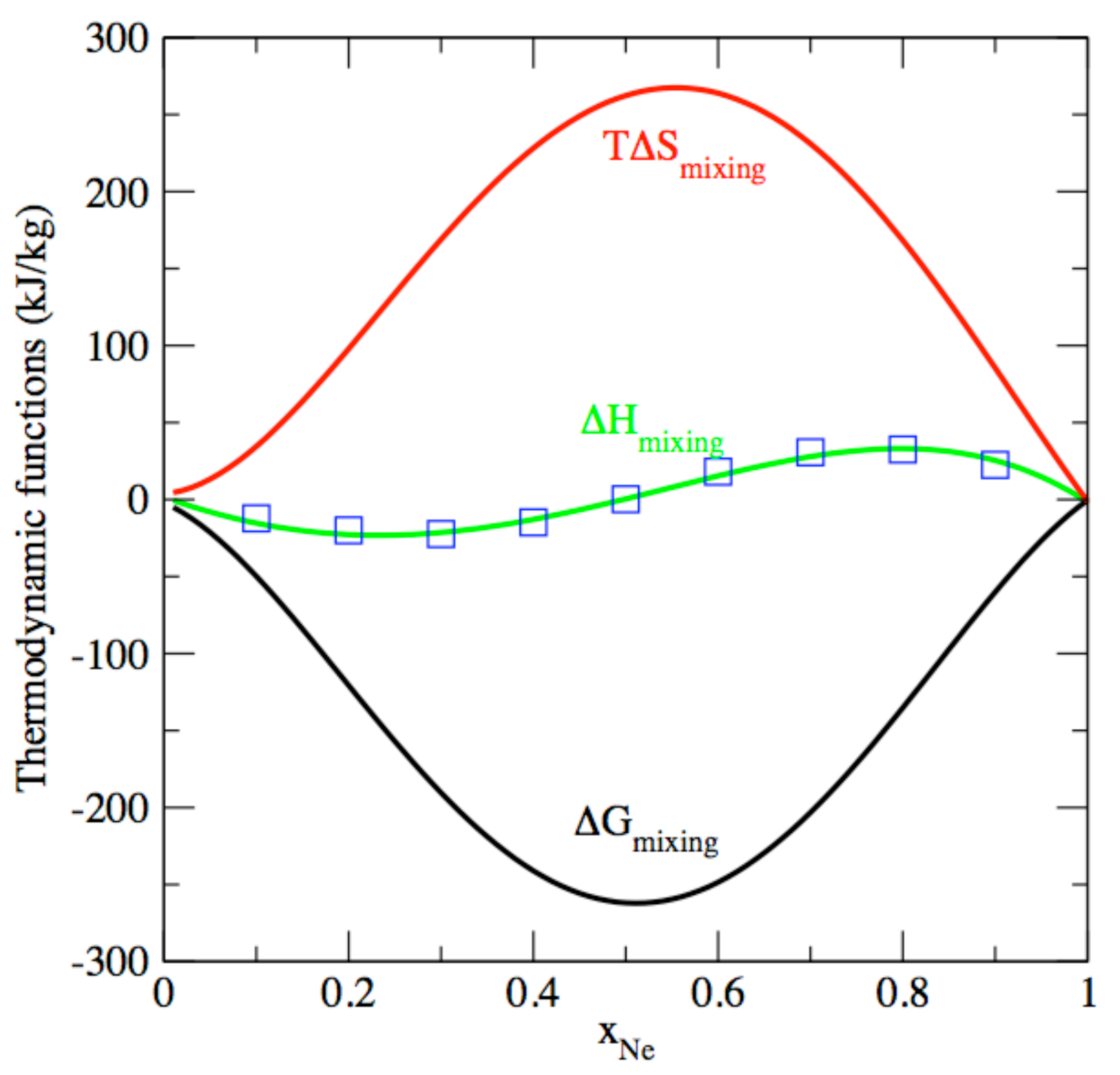}(a)
\includegraphics*[width=10cm]{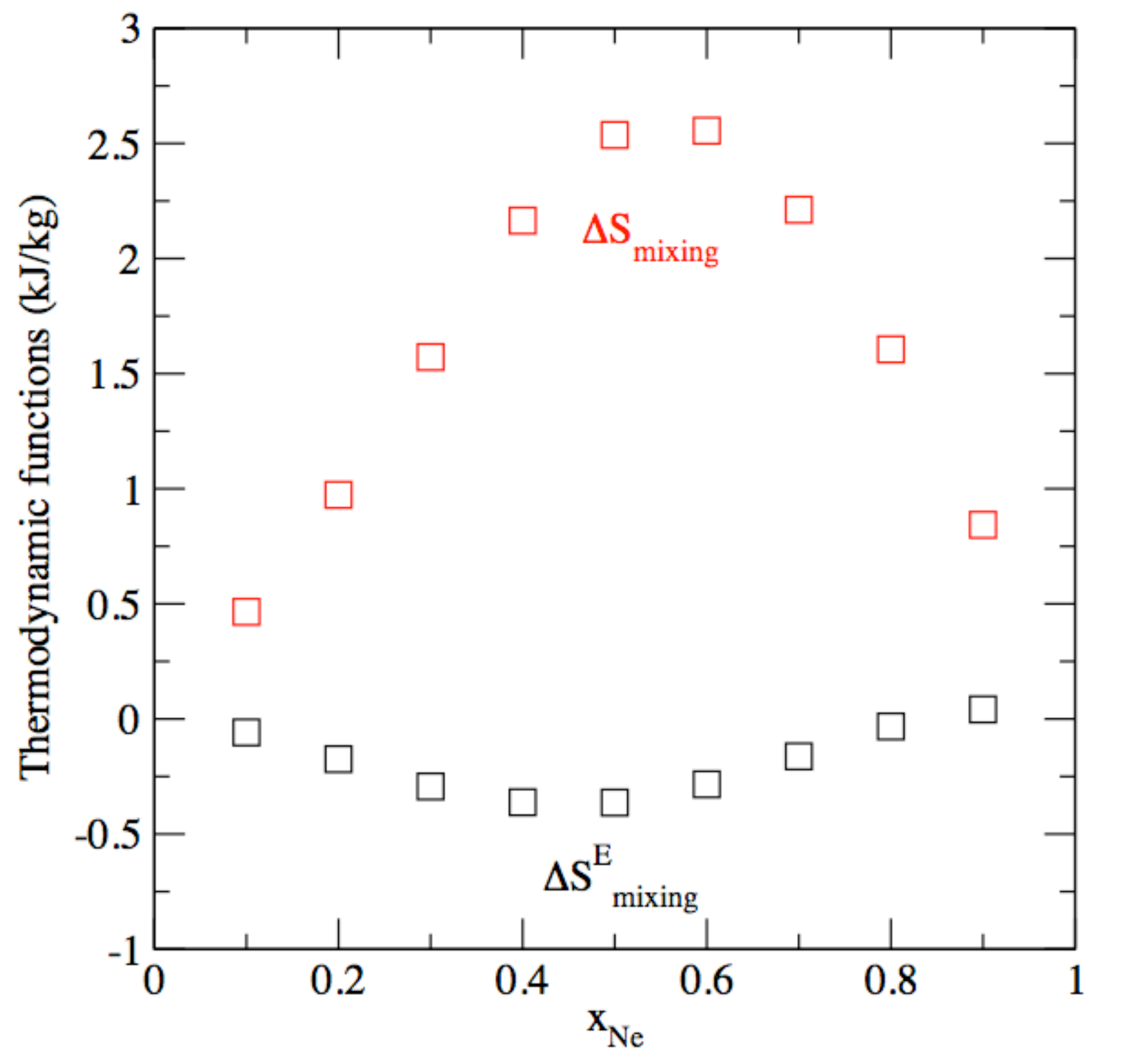}(b)
\end{center}
\caption{Variation of the thermodynamic properties of mixing, with in (a), $(\mu_{Ne}, \mu_{Ar}, P, R)$ simulation results for $\Delta H_{mix}$ (green triangles), $T\Delta S_{mix}$ (red squares) and $\Delta G_{mix}$ (black circles) as a function of $x_{Ne}$. $(N_{Ne}, N_{Ar}, P, T)$ simulation results for $\Delta H_{mix}$ are shown as blue squares. In (b), the excess entropy of mixing $\Delta S^E_{mix}$ (black squares) is compared to the entropy of mixing $T\Delta S_{mix}$ (red squares).}
\label{Fig6}
\end{figure}

We report in Table~\ref{Tab5} the results obtained from $(\mu_{Ne}, \mu_{Ar}, P, R)$ simulations for the mixtures properties, as well as the thermodynamic properties of mixing, for different mole fractions in Ne. We fit the simulation results to obtain the following equations for the thermodynamic properties of mixing
\begin{equation}
\begin{array}{ccc}
\Delta \bar{H}_{mix}~(kJ/kg)=1.664-218.26x_{Ne}+478.90x^2_{Ne}+74.72x^3_{Ne}-338.65x^4_{Ne}\\
\Delta \bar{S}_{mix}~(kJ/kg/K)=0.034+0.408x_{Ne}+28.545x^2_{Ne}-51.101x^3_{Ne}+22.103x^4_{Ne}\\
\Delta \bar{G}_{mix}~(kJ/kg)=-2.119-263.46x_{Ne}-2683.3x^2_{Ne}+5735.7x^3_{Ne}-2787.2x^4_{Ne}\\
\end{array}
\label{NeAr}
\end{equation}
We also plot in Fig.~\ref{Fig6} the dependence of the thermodynamic properties of mixing as a function of the mole fraction in Ne. We provide a comparison between the enthalpy of mixing predicted by $(N, P, T)$ simulations for the corresponding mole fractions at $T=110.8$~K and $P=445$~bar. The results show that there is a very good agreement between the simulations results obtained from both sets of simulations, thereby providing another validation for the $(\mu_{Ne}, \mu_{Ar}, P, R)$ simulation method. $(\mu_{Ne}, \mu_{Ar}, P, R)$ simulations have the advantage of also providing the other thermodynamic properties of mixing, {\it i.e.} the entropy of mixing and the Gibbs free energy of mixing. 

Fig.~\ref{Fig6}(a) shows that the entropy of mixing exhibits a maximum for an equimolar mixture, while the enthalpy of mixing remains close to 0~kJ/kg and only increases slightly with the Ne mole fraction. This results in the presence of a minimum for the Gibbs free energy of mixing for a mole fraction of 0.5. Overall, the plot shown in Fig.~\ref{Fig6} is close to what is expected for an ideal binary mixture. Indeed, ideal mixtures exhibit an enthalpy of mixing of 0~kJ/kg and two terms of the same magnitude, and opposite signs, for the Gibbs free energy of mixing and for the product of temperature by the entropy of mixing. Furthermore, for ideal mixtures, the Gibbs free energy of mixing reaches a minimum for a mole fraction of 0.5, while the entropy of mixing reaches a maximum for a mole fraction of 0.5. Since we have a mixture of two rare gases, it is reasonable to observe here a behavior that is qualitatively similar to that of ideal mixtures. This confirms the ability of the $(\mu_{Ne},\mu_{Ar},P,R)$ simulations to yield the thermodynamic properties of mixing. We also examine in Fig.~\ref{Fig6}(b) the results obtained for the excess entropy of mixing, defined as the entropy of the mixture minus the ideal gas entropy of the two components for the mixture. The excess entropy of mixing is small, less than 0.5 kJ/kg/K in absolute value, when compared to the entropy of mixing. This means that intermolecular interactions contribute very little to the entropy of mixing of Ne-Ar and that the main contribution to the entropy of mixing is thus combinatorial. This is in line with the results obtained for the enthalpy of mixing, which show that the magnitude of this term is small.

\subsection{The Cu-Ag system}

\subsubsection{Single-component systems}

We now turn to the study of metallic systems, and focus on the example of the Cu-Ag mixture. We start by examining the results for single-component systems of Cu and Ag. Table~\ref{Tab6} and Table~\ref{Tab7} show the results obtained for the two metals along the $P=1$~bar isobar. We first comment on the results for Cu (Table~\ref{Tab6}), and find that the specific volume increases as the chemical potential decreases. This leads to fewer, and weaker, interactions between Cu atoms and, in turn, to a decrease in enthalpy. This is confirmed by the increase in temperature and entropy, which shows that the fluid is less and less organized as the chemical potential decreases. 

\begin{table}[ht]
\centering
\begin{tabular}{ccccccc}
\hline\hline
$\mu$~(kJ/kg)~&~$<T>$~(K)~&~$\bar{V}$~(cm$^3$/g)~&~$\bar{H}$~(kJ/kg)~&~$\bar{R}$~(kJ/kg)~&~$\bar{S}$~(kJ/kg/K)\\
\hline
-6300 & 1306.1 & 0.129 & -4585.23 & 1714.8 & 1.313  \\
-6400 & 1385.6 & 0.131 & -4541.45 & 1858.6 & 1.341  \\
-6500 & 1455.7 & 0.132 & -4506.60 & 1993.4 & 1.369  \\
-6600 & 1535.2 & 0.133 & -4466.60 & 2132.8 & 1.389  \\
-6700 & 1602.2 & 0.135 & -4431.64 & 2268.6 & 1.416  \\
-6800 & 1680.0 & 0.136 & -4392.88 & 2406.9 & 1.433  \\
-6900 & 1737.5 & 0.137 & -4363.18 & 2537.5 & 1.460  \\
-7000 & 1809.7 & 0.138 & -4329.20 & 2670.7 & 1.476  \\
-7100 & 1873.9 & 0.139 & -4299.75 & 2800.0 & 1.494  \\
-7200 & 1937.5 & 0.141 & -4263.77 & 2936.3 & 1.515  \\
\hline\hline
\end{tabular}
\caption{Copper: $(\mu,P,R)$ simulation results along the $P=1$~bar isobar, with $R/k_B=8 \times 10^5$~K.}
\label{Tab6}
\end{table}

\begin{figure}
\begin{center}
\includegraphics*[width=12cm]{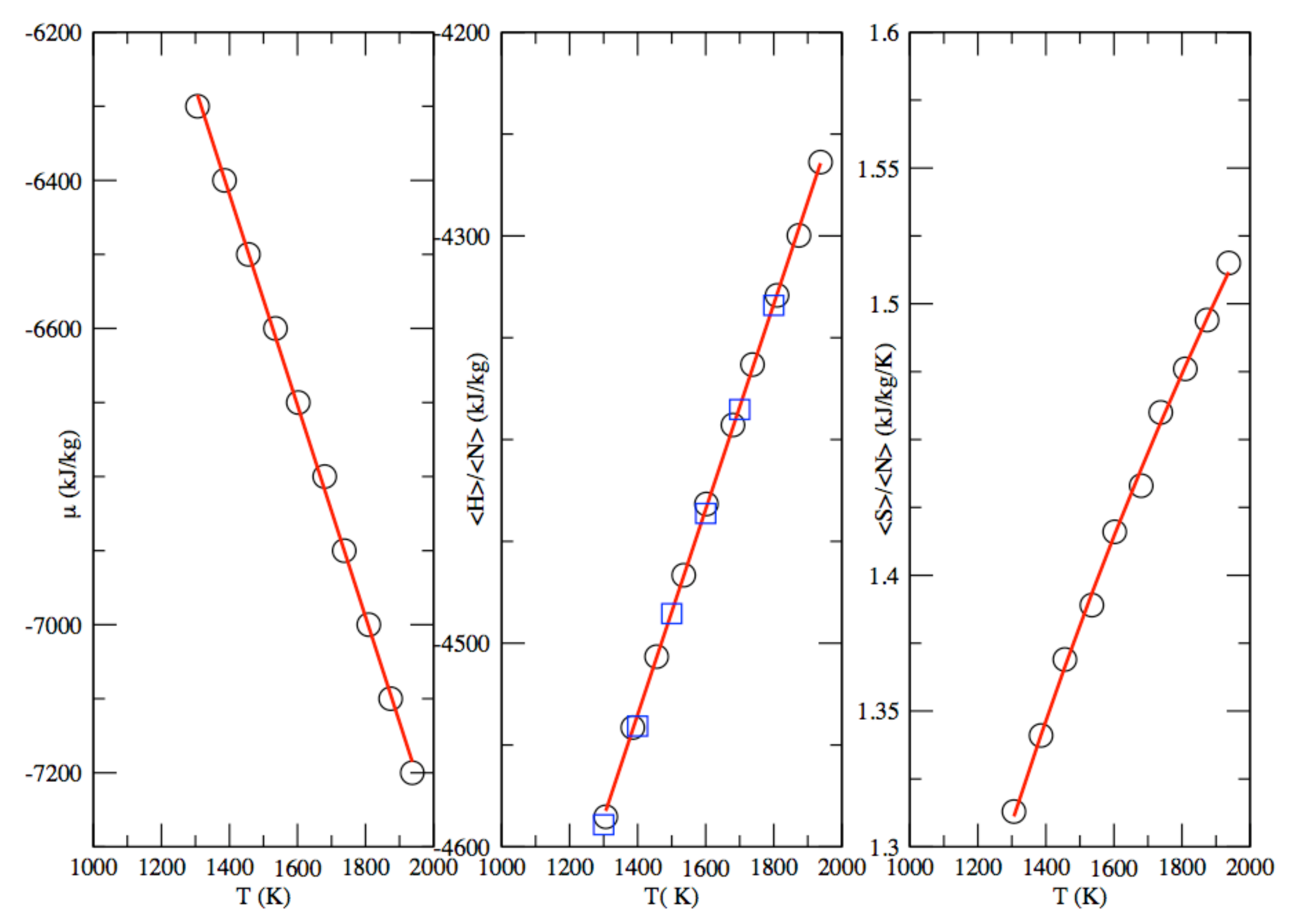}
\end{center}
\caption{Copper along the $P=1$~bar isobar. (a) Chemical potential against temperature, (b) Enthalpy against temperature, and (c) Entropy against temperature. $(\mu,P,R)$ simulation results are shown as black circles, $(N,P,T)$ simulation results are shown as blue squares, and fits to the $(\mu,P,R)$ simulation results are shown as red lines.}
\label{Fig7}
\end{figure}

We show in Fig.~\ref{Fig7}, the variation of the chemical potential, enthalpy, and entropy as a function of temperature. Fig.~\ref{Fig7} demonstrates that there is a good agreement between results obtained with $(\mu,P,R)$ simulations and results from Monte Carlo simulations in the isothermal-isobaric ensemble (see middle panel of Fig~\ref{Fig7}). In line with the noble gases systems, we fit the simulation results and obtain a linear fit for $\mu(T)$. We obtain $\mu(T)$~(kJ/kg)~$=-4422.0 - 1.426 \times T$, yielding an estimate for the average entropy over the temperature interval of $1.426$~kJ/kg/K. For enthalpy and entropy, we obtain the following fits
\begin{equation}
\begin{array}{c}
\bar{H}~(kJ/kg)~=-5240.6 + 0.504 \times T\\
\bar{S}~(kJ/kg/K)=-2.337 + 0.508 \ln T\\
\end{array}
\label{FitsCu}
\end{equation}

\begin{table}[ht]
\centering
\begin{tabular}{cccccc}
\hline\hline
$\mu$~(kJ/kg)~&~$<T>$~(K)~&~$\bar{V}$~(cm$^3$/g)~&~$\bar{H}$~(kJ/kg)~&~$\bar{R}$~(kJ/kg)~&~$\bar{S}$~(kJ/kg/K)\\
\hline
-3200 & 1133.4 & 0.112 & -2258.59 & 941.4 & 0.831  \\
-3300 & 1218.5 & 0.113 & -2228.54 & 1071.5 & 0.879  \\
-3400 & 1359.7 & 0.116 & -2182.48 & 1218.0 & 0.896  \\
-3500 & 1456.4 & 0.118 & -2151.29 & 1348.6 & 0.926  \\
-3600 & 1573.9 & 0.120 & -2116.86 & 1483.2 & 0.942  \\
-3700 & 1680.2 & 0.122 & -2084.49 & 1615.7 & 0.962  \\
-3800 & 1774.9 & 0.124 & -2053.67 & 1746.2 & 0.984  \\
-3900 & 1892.8 & 0.126 & -2020.79 & 1878.0 & 0.992  \\
-4000 & 1993.5 & 0.128 & -1993.93 & 2006.8 & 1.007  \\
\hline\hline
\end{tabular}
\caption{Silver: $(\mu,P,R)$ simulation results along the $P=1$~bar isobar, with $R/k_B=8 \times 10^5$~K.}
\label{Tab7}
\end{table}

\begin{figure}
\begin{center}
\includegraphics*[width=12cm]{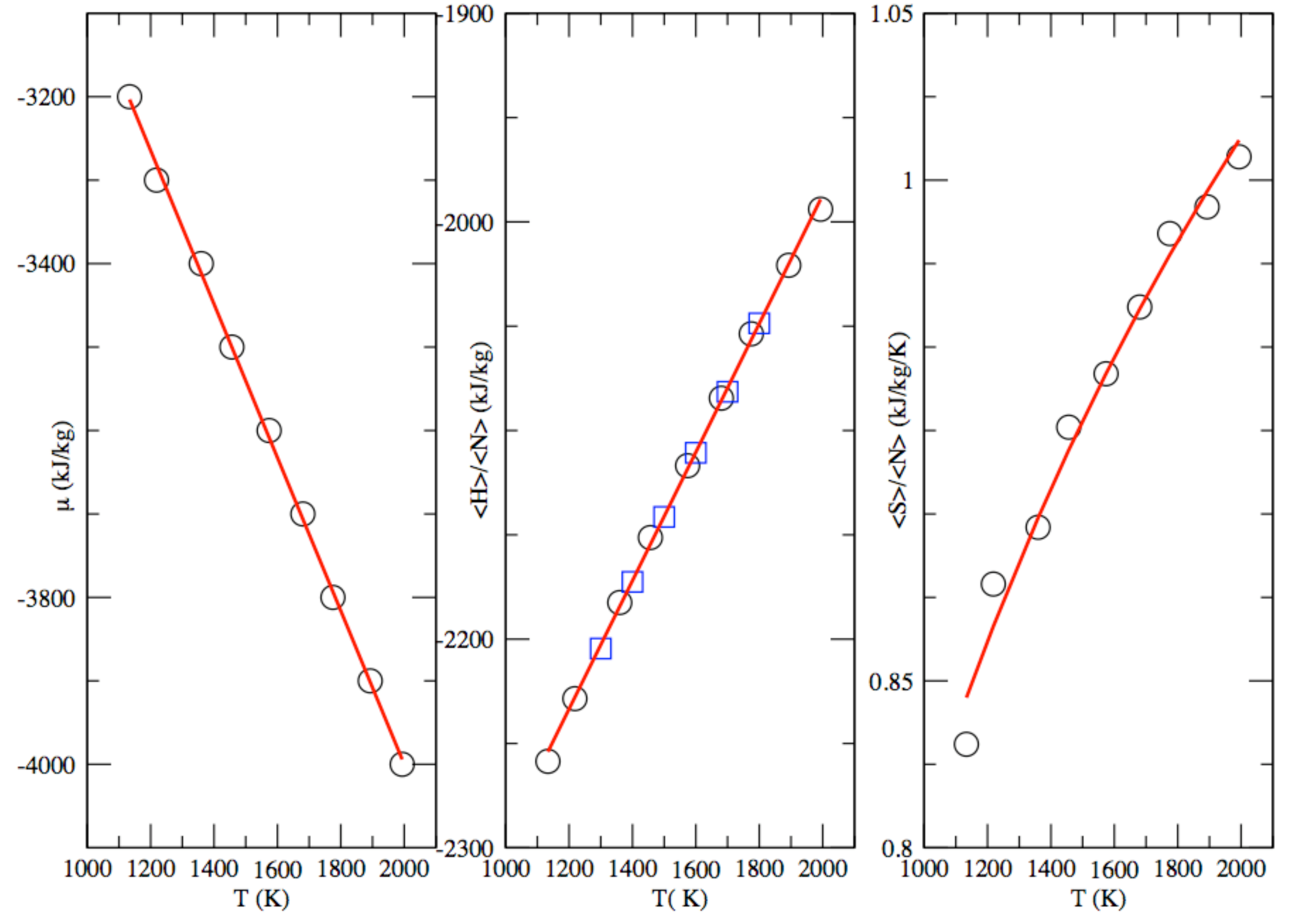}
\end{center}
\caption{Silver along the $P=1$~bar isobar. (Left panel) Chemical potential against temperature, (Middle panel) Enthalpy against temperature, and (Right panel) Entropy against temperature. $(\mu,P,R)$ simulation results are shown as black circles, $(N,P,T)$ simulation results are shown as blue squares, and fits to the $(\mu,P,R)$ simulation results are shown as red lines.}
\label{Fig8}
\end{figure}

Next, we turn to the results obtained for Ag and provide in Table~\ref{Tab7} the results obtained from $(\mu,P,R)$ simulations. We then compare the results for the specific volume to reference data~\cite{assael2012reference} over the $1235$~K-$1600$~K range and find a good agreement. For instance, at $T=1359.7$~K, the reference data is of $\bar{V}^{ref}=0.109$~cm$^3$/g to be compared to $0.116$~cm$^3$/g predicted by the simulation, and at $T=1573.9$~K, we have $\bar{V}^{ref}=0.111$~cm$^3$/g to be compared to $0.120$~cm$^3$/g for the simulation. We show in Fig.~\ref{Fig8} plots of the chemical potential, enthalpy, and entropy against temperature. As for copper, we observe a good agreement between the enthalpy predicted by $(\mu,P,R)$ simulations and that obtained with Monte Carlo $(N,P,T)$ simulations (see Fig~\ref{Fig8}). We then carry out the same analysis as above and obtain the following linear fit for $\mu(T)$~(kJ/kg)~$=-2161.0 - 0.920 \times T$, which provides an estimate for the average entropy over the temperature interval of $0.920$~kJ/kg/K. For enthalpy and entropy, we obtain the following fits
\begin{equation}
\begin{array}{c}
\bar{H}~(kJ/kg)=-2603.0 + 0.308 \times T\\
\bar{S}~(kJ/kg/K)=-1.235 + 0.296 \ln (T)\\
\end{array}
\label{FitsAg}
\end{equation}

\subsubsection{Thermodynamic properties of the Cu-Ag mixture}

\begin{figure}
\begin{center}
\includegraphics*[width=12cm]{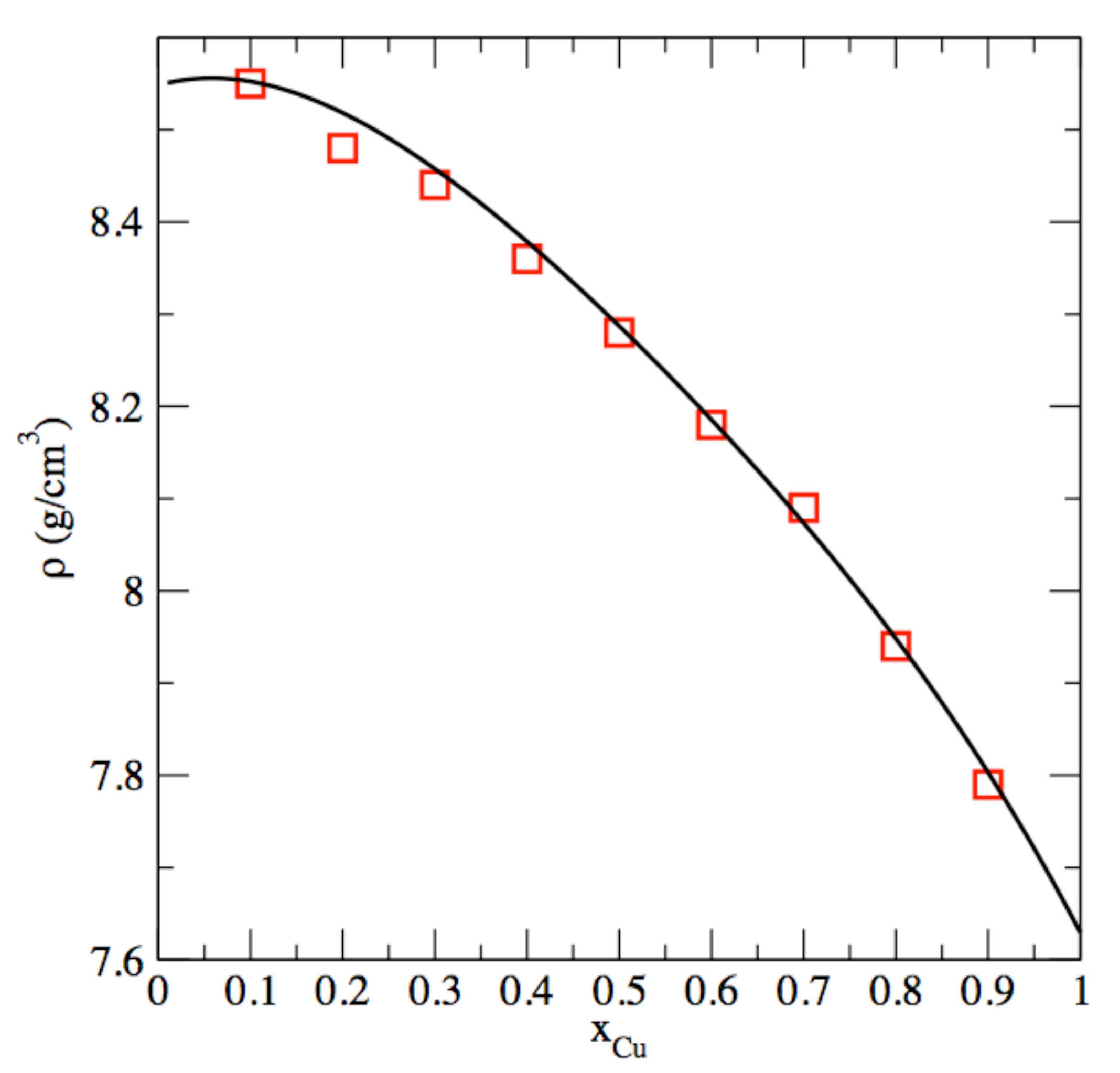}
\end{center}
\caption{Density of the Cu-Ag mixture at $1$~bar and $1400$~K. The black line is a plot of Eq.~\ref{fitrhomCuAg}, obtained from $(\mu_{Cu},\mu_{Ag},P,R)$ simulation results, while the open red squares are $(N_{Cu},N_{Ag},P,T)$ simulation results.}
\label{Fig9}
\end{figure}

We then examine the properties for the Cu-Ag mixture. To this end, we carry out $(\mu_{Cu},\mu_{Ag},P,R)$ simulations for the mixture for conditions corresponding to $P=1$~bar and an average temperature of $1400~\pm20$~K, and fit the simulation results to determine a series of equations modeling the properties of the mixture as a function of the mole fraction in copper. We start with the density of the system, and obtain the following equation from the simulation results
\begin{equation}
\rho~(g/cm^3)=8.548+0.278x_{Cu}-2.614x^2_{Cu}+2.639x^3_{Cu}-1.221x^4_{Cu}
\label{fitrhomCuAg}
\end{equation}
and test the fit against the results obtained from $(N_{Cu},N_{Ag},P,T)$ simulation results. The results are shown in Fig.~\ref{Fig9}. They are found to be in very good agreement with one another over the entire range of compositions, thereby establishing that Eq.~\ref{fitrhomCuAg} provides an accurate model for the density of the Cu-Ag mixture.

\begin{figure}
\begin{center}
\includegraphics*[width=12cm]{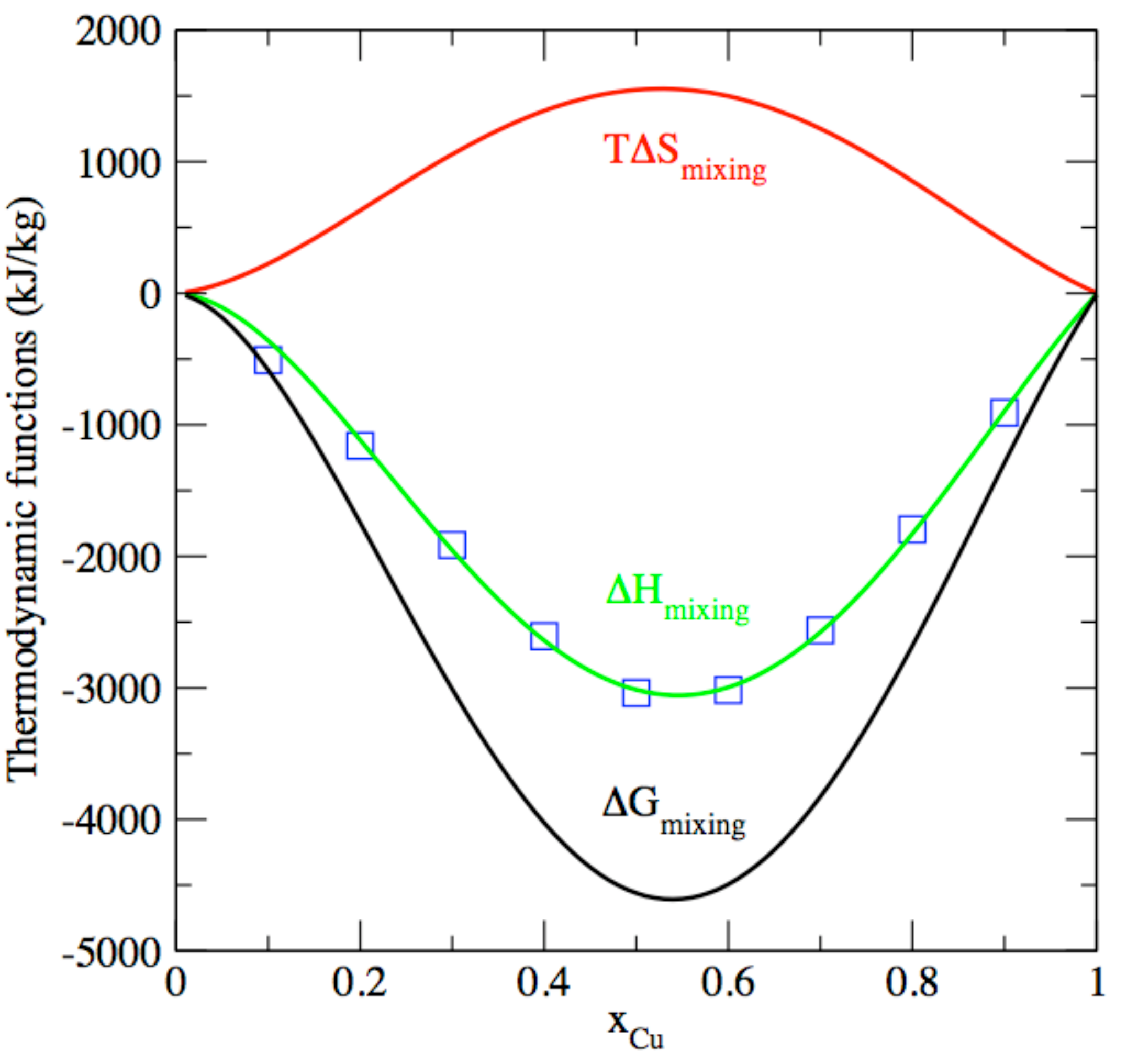}
\end{center}
\caption{Thermodynamic properties of mixing for the Cu-Ag mixture at $1$~bar and $1400$~K. The black line is a plot of Eq.~\ref{fitCuAg}, obtained from $(\mu_{Cu}, \mu_{Ag}, P, R)$ simulation results, while the open blue squares are $(N_{Cu}, N_{Ag}, P, T)$ simulation results.}
\label{Fig10}
\end{figure}

Next, we determine the corresponding equations for the thermodynamic properties of mixing using Eq.~\ref{Mixprop}. To this end, we take the $(\mu_{Cu}, \mu_{Ag}, P, R)$ simulation results for the mixture, {\it i.e.}, $<\bar{H}_m>$ and $<\bar{S}_m>$ and calculate $<\bar{G}_m>=<\bar{H}_m>-T<\bar{S}_m>$. Then, we subtract the value for $\bar{H}$ and $\bar{S}$ obtained for the single-component systems under the same conditions of pressure and temperature. These are provided by Eq.~\ref{FitsCu} for Cu as $\bar{H}_{Cu}=-4535.0$~kJ/kg and $\bar{S}_{Cu}=1.343$~kJ/kg/K and by Eq.~\ref{FitsAg} for Ag as $\bar{H}_{Ag}=-2171.8$~kJ/kg and $\bar{S}_{Ag}=0.909$~kJ/kg/K. We obtain the following equations for the thermodynamic properties of mixing
\begin{equation}
\begin{array}{c}
\Delta \bar{H}_{mix}~(kJ/kg)=0.532-403.75x_{Cu}-38825x^2_{Cu}+71068x^3_{Cu}-31847x^4_{Cu}\\
\Delta \bar{S}_{mix}~(kJ/kg/K)=0.504x_{Cu}+13.503x^2_{Cu}-26.369x^3_{Cu}+12.365x^4_{Cu}\\
\Delta \bar{G}_{mix}~(kJ/kg)=0.982-1109.8x_{Cu}-57729x^2_{Cu}+107984x^3_{Cu}-49158x^4_{Cu}\\
\end{array}
\label{fitCuAg}
\end{equation}

We plot in Fig.~\ref{Fig10} the resulting fits, as well as a comparison with results obtained from a series of $(N_{Cu}, N_{Ag}, P, T)$ simulations as we vary the composition of the mixture at $P=1$~bar and $T=1400$~K. As shown in Fig.~\ref{Fig10}, there is an excellent agreement for $\Delta \bar{H}_{m}$ between the fits to the $(\mu_{Cu}, \mu_{Ag}, P, R)$ results of Eq.~\ref{fitCuAg} and the $(N_{Cu}, N_{Ag}, P, T)$ simulation results. This validates the extension of the method proposed in this work for mixtures, and its applicability to many-body force fields. Furthermore, the results obtained from both sets of simulation indicate that the behavior of the Cu-Ag mixture departs from that observed for ideal mixtures. Specifically, the enthalpy of mixing takes values that are of the same order as the other two terms, $\Delta \bar{G}_m$ and $T \Delta \bar{S}_m$. In other words, the enthalpy of mixing is not negligible any longer as one would expect for an ideal mixture. Furthermore, the minima for $\Delta \bar{H}_m$ and $\Delta \bar{G}_m$, and the maximum for $T \Delta \bar{S}_m$, are reached for a mole fraction in Ne that is now about 0.55, and not 0.5 any longer as for an ideal mixture. This departure can be interpreted as stemming from the strong cohesive interactions, and of dramatically different magnitudes, that take place between Cu and Ag atoms. It also illustrates one of the key advantages of the $(\mu_{Cu}, \mu_{Ag}, P, R)$ method, as it provides direct access to all three quantities, $G$, $H$ and $S$, during a simulation run.

\section{Conclusions}
In this work, we extend the adiabatic formalism to multicomponent systems and, more specifically, to the adiabatic grand-isobaric ensemble. Then, we develop an implementation of simulations in the adiabatic grand-isobaric ensemble within a Monte Carlo framework and apply the new approach to binary mixtures of noble gases and of metals. We show that this method has two very significant advantages. First, we now have direct access to the entropy of the mixture through the relation $R=ST$. Second, the calculation of the pressure through the virial relation is not required here, since pressure is an input parameter in simulations in the adiabatic grand-isobaric ensemble. This alleviates the need for the computation of pressure, and its increased complexity when many-body terms are included. This new approach is thus particularly well suited for the determination of the entropy of mixing, an issue that has drawn considerable interest since the dawn of statistical mechanics, and of the other thermodynamic properties of mixing, including the enthalpy of mixing and the Gibbs free energy of mixing. We assess the accuracy of the method through comparisons with the available experimental data on mixtures of ideal gases, and with results obtained from conventional simulations performed in the isothermal-isobaric ensemble. This new approach allows us to recover the ideal behavior expected for mixtures of noble gases. Furthermore, simulations in the adiabatic grand-isobaric ensemble sheds light on the departure from the ideal behavior observed in binary metallic mixtures. Specifically, in the case of the Cu-Ag mixture, we observe a shift in the maximum for the entropy of mixing towards a greater Cu content than the ideal value of 0.5, and a strong contribution of the enthalpy of mixing to the Gibbs free energy of mixing. This results also shows the versatility of the adiabatic grand-isobaric approach, and its applicability to systems modeled with many-body force fields. The extension of the method to molecular fluids is currently under way.

\begin{acknowledgments}
Partial funding for this research was provided by NSF through award CHE-1955403. This work used the Extreme Science and Engineering Discovery Environment (XSEDE)~\cite{xsede}, which is supported by National Science Foundation grant number ACI-1548562, and used the Open Science Grid through allocation TG-CHE200063.
\end{acknowledgments}

\vspace{1 cm}

{\bf Data availability}
\vspace{0.5 cm}

The data that support the findings of this study are available from the corresponding author upon reasonable request.

\bibliography{JCP_Binary}

\end{document}